\documentclass[useAMSD,usenatbib]{mn2e}
\pdfoutput=1

\usepackage{aas_macros}
\usepackage{amsmath}
\usepackage{amssymb}
\usepackage{booktabs}
\usepackage{color}
\usepackage{relsize}
\usepackage{capt-of}
\usepackage[english]{babel}
\usepackage{epsfig}
\usepackage{float}
\usepackage[T1]{fontenc}
\usepackage{graphicx}
\usepackage[utf8]{inputenc}
\usepackage{mathrsfs}
\usepackage{subfig}
\usepackage{times}
\usepackage{hyperref}

\hypersetup{pagebackref=true, colorlinks=true,  linkcolor=red, citecolor=blue, urlcolor=blue, bookmarks=false}

\title[Tracing the cosmic growth of SMBHs with \textit{Herschel}]{Tracing the cosmic growth of super massive black holes to \textit{z}$\sim$3 with \textit{Herschel} 
\thanks{{\textit{Herschel} is an ESA space observatory with science instruments provided by European-led Principal Investigator consortia and with important participation from NASA.} }
}

\author[I.~Delvecchio et al.]
{I.~Delvecchio$^{1}$\thanks{E-mail: ivan.delvecchio@unibo.it},
C. Gruppioni$^{2}$,
F. Pozzi$^{1}$,
S. Berta$^{3}$,
G. Zamorani$^{2}$,
A. Cimatti$^{1}$,
\newauthor 
D. Lutz$^{3}$,
D. Scott$^{4}$,
C. Vignali$^{1}$, 
G. Cresci$^{5}$,
A. Feltre$^{6,7}$,
A. Cooray$^{8,9}$,
M. Vaccari$^{10}$,
\newauthor 
J. Fritz$^{11}$,
E. Le Floc'h$^{12}$,
B. Magnelli$^{13}$,
P. Popesso$^{3}$,
S. Oliver$^{14}$,
J. Bock$^{9,15}$,
M. Carollo$^{16}$,
\newauthor 
T. Contini$^{17}$,
O. Le F\'evre$^{18}$,
S. Lilly$^{16}$,
V. Mainieri$^{7}$,
A. Renzini$^{19}$ and M. Scodeggio$^{20}$
\\ \\
$^{1}$ Dipartimento di Fisica e Astronomia, Universit\`a di Bologna, via Ranzani 1, I-40127 Bologna, Italy.\\
$^{2}$ INAF - Osservatorio Astronomico di Bologna, via Ranzani 1, I-40127 Bologna, Italy.\\
$^{3}$ Max-Planck-Institut f\"or Extraterrestrische Physik (MPE), Postfach 1312, D-85741 Garching, Germany.\\
$^{4}$ Department of Physics and Astronomy, University of British Columbia, 6224 Agricultural Road, Vancouver, BC V6T 1Z1, Canada. \\
$^{5}$ INAF - Osservatorio Astrofisico di Arcetri, Largo E. Fermi 5, 50125 Firenze, Italy. \\
$^{6}$ Dipartimento di Fisica e Astronomia, Universit\`a di Padova, vicolo Osservatorio, 3, 35122 Padova, Italy.\\
$^{7}$ ESO, Karl-Schwarzschild-Str. 2, 85748 Garching bei M\"unchen, Germany.\\
$^{8}$ Dept. of Physics \& Astronomy, University of California, Irvine, CA 92697, USA \\
$^{9}$ Jet Propulsion Laboratory, 4800 Oak Grove Drive, Pasadena, CA 91109, USA \\
$^{10}$ Astrophysics Group, Department of Physics, University of Western Cape, Bellville 7535, Cape Town, South Africa. \\
$^{11}$ Sterrenkundig Observatorium, Vakgroep Fysica en Sterrenkunde Universeit, Gent, Krijgslaan 281, S9 9000 Gent, Belgium.\\
$^{12}$ CEA-Saclay, Service d'Astrophysique, F-91191 Gif-sur-Yvette, France. \\
$^{13}$ Argelander Institute for Astronomy, Bonn University, Auf dem H\"ugel 71, D-53121 Bonn, Germany. \\
$^{14}$ Astronomy Centre, Department of Physics and Astronomy, University of Sussex, Brighton BN1 9QH, UK. \\
$^{15}$ California Institute of Technology, 1200 E. California Blvd., Pasadena, CA 91125, USA \\
$^{16}$ Institute of Astronomy, Swiss Federal Institute of Technology (ETH H\"onggerberg), CH-8093, Z\"urich, Switzerland. \\
$^{17}$ Institut de Recherche en Astrophysique et Planétologie, CNRS, Université de Toulouse, 14 avenue E. Belin 31400 Toulouse, France. \\
$^{18}$ Laboratoire d'Astrophysique de Marseille, CNRS-Universit\`e de Provence, rue Fr\`ed\`eric Joliot-Curie 38, 13388 Marseille Cedex 13, France. \\
$^{19}$ INAF Osservatorio Astronomico di Padova, vicolo dell’Osservatorio 5, I-35122 Padova, Italy. \\
$^{20}$ INAF – IASF Milano, via Bassini 15, 20133 Milano, Italy. }

\begin{document}

\date{Accepted 2014 January 16. Received 2013 December 30; in original form 2013 November 14}

\pagerange{\pageref{firstpage}--\pageref{lastpage}} \pubyear{2014}

\maketitle

\label{firstpage}

\begin{abstract}
We study a sample of \textit{Herschel}-PACS selected galaxies within the GOODS-South and the COSMOS fields in the framework of the PACS Evolutionary Probe (PEP) project. Starting from the rich multi-wavelength photometric data-sets available in both fields, we perform a broad-band Spectral Energy Distribution (SED) decomposition to disentangle the possible active galactic nucleus (AGN) contribution from that related to the host galaxy. We find that 37 per cent of the \textit{Herschel}-selected sample shows signatures of nuclear activity at the 99 per cent confidence level. The probability to reveal AGN activity increases for bright ($L_{\rm 1-1000} > 10^{11} \rm L_{\odot}$) star-forming galaxies at $z>0.3$, becoming about 80 per cent for the brightest ($L_{\rm 1-1000} > 10^{12} \rm L_{\odot}$) infrared (IR) galaxies at $z \geq 1$. Finally, we reconstruct the AGN bolometric luminosity function and the super-massive black hole growth rate across cosmic time up to $z \sim 3$ from a Far-Infrared (FIR) 
perspective. This work shows general agreement with most of the panchromatic estimates from the literature, with the global black hole growth peaking at $z \sim 2$ and reproducing the observed local black hole mass density with consistent values of the radiative efficiency $\epsilon_{\rm rad}$ ($\sim$0.07).
\end{abstract}


\section{Introduction} \label{intro}

For more than five decades, since the discovery of quasars, there has been an increasing interest in understanding such extremely bright objects. Though the detailed mechanisms that are the source of the powerful nuclear activity are still debated (\citealt{Rees78}; \citealt{Lodato+07}; \citealt{Devecchi+09}; \citealt{Volonteri+10}; \citealt{Ball+11}; \citealt{Natarajan11}), it is believed that it is mainly due to mass accretion onto a super massive black hole (SMBH, \citealt{Salpeter64}; \citealt{Lynden-Bell69}; \citealt{Shakura+73}; \citealt{Soltan82}; \citealt{Rees84}), allowing them to be revealed as active galactic nuclei (AGN). In the last few years several studies have highlighted the importance of AGN in the framework of galaxy formation and evolution. 

Indeed, many pioneering studies have shown the presence of a correlation between the BH mass and other physical properties of its host: galaxy bulge stellar mass, luminosity, velocity dispersion (\citealt{Magorrian+98}; \citealt{Ferrarese+00}; \citealt{Gebhardt+00}; \citealt{Tremaine+02}; \citealt{Marconi+03}; \citealt{Gultekin+09}) and dark matter halo mass (\citealt{Ferrarese+02}). 

Observations have suggested a tight AGN/starburst connection through the ``downsizing'' scenario, according to which most massive BHs accrete earlier, faster and with more extreme accretion rates than their lower mass counterparts (\citealt{Ueda+03}; \citealt{LeFloc'h+05}; \citealt{Perez-Gonzalez+05}; \citealt{Hasinger+05}; \citealt{Richards+06}). A similar anti-hierarchical growth seems to be mirrored in the evolution of the cosmic star formation density (SFD, \citealt{Lilly+96}; \citealt{Madau+96}; \citealt{Cowie+96}, \citeyear{Cowie+97}; \citealt{Bell+05}; \citealt{Juneau+05}; \citealt{Bundy+06}; \citealt{Wall+08}; \citealt{Gruppioni+13}), hinting at a deep interplay among the AGN and its host.

Furthermore, semi-analytical models have shown that stellar processes on their own struggle to reproduce the currently observed galaxy population properties, without adding a further source of energy (\citealt{Benson+03}; \citealt{Bower+06}; \citealt{Schawinski+06}; \citealt{Gabor+11}). This required injection of heat is believed to come from the central active black hole, which seems to affect the surrounding galactic environment. Thorough theoretical studies have well established that nuclear accretion processes could impact the star formation history (SFH) of the whole galaxy, making unavailable a fraction of the cold gas required to drive star formation, either by removing or by heating it up (\citealt{Springel+05}; \citealt{Croton+06}; \citealt{Hopkins+06}; \citealt{Menci+06}). It is clear that a comprehensive study of the AGN accretion history is crucial to shed light on the evolution of galaxies down to the present epoch.

So far, different accretion scenarios have been proposed: major mergers of gas-rich galaxies (\citealt{Sanders+88}; \citealt{DiMatteo+05}; \citealt{Hopkins+06}; \citealt{Menci+08}; \citealt{Hopkins+10}); gradually varying processes (\citealt{Kormendy+04}), which in turn include internal secular evolution (i.e. large-scale disc instabilities and bars) and external secular evolution (galaxy interactions) and inflow of recycled cold gas from old stellar populations (\citealt{Ciotti+07}; \citealt{Ciotti+10}). Though an accurate determination of the AGN fraction related to each single scenario has not yet been derived, it is currently believed that most luminous AGN ($L_{\rm AGN} > 10^{44}$ erg s$^{-1}$) are triggered by major merger events (\citealt{Netzer09}; \citealt{Santini+12}), while their fainter counterparts are fueled through secular processes (\citealt{Shao+10}; \citealt{Lutz+10}; \citealt{Rosario+12}).

Recent attempts have been made to trace the SMBH growth over cosmic time (e.g. \citealt{Marconi+04}; \citealt{Shankar+04}; \citealt{Hasinger+05}; \citealt{Hopkins+06}; \citealt{Hopkins+07}; \citealt{Merloni+08}; \citealt{Aird+10}; \citealt{Fanidakis+12}; \citealt{Mullaney+12}), supporting the connection between black hole accretion density (BHAD) and SFD up to $z \simeq$ 1--1.5. In particular, \citet{Mullaney+12} suggest that the ratio SFD/BHAD $\sim$ 1--2$~ \times~10^3$ in the redshift range $0.5<z<2.5$. In agreement with observations, semi-analytical models and hydro-dynamical simulations currently hint at an overall decrease of the BHAD from $z \simeq 2$ down to the present day. At higher redshift the situation is more uncertain, since direct diagnostics for black hole mass estimates (e.g. reverberation, time variability, etc.) are no longer available, and indirect measurements need to be extrapolated up to these redshifts. 

To date the AGN obscured growth is largely uncertain beyond $z \simeq$ 1--2, since even the deepest X-ray surveys suffer from a decreasing completeness with increasing redshift and/or increasing obscuration (\citealt{Brandt+05}). While the dependence of $N_{\rm H}$ (hydrogen column density) on the intrinsic AGN luminosity is currently well established (\citealt{Lawrence91}; \citealt{Simpson05}; \citealt{Ueda+03}; \citealt{Steffen+03}; \citealt{Barger+05}; \citealt{Sazonov+07}; \citealt{Lusso+13}), the evolution with redshift is still debated: on the one hand, a slight evolution with redshift is suggested by some analyses (\citealt{LaFranca+05}; \citealt{Treister+06}; \citealt{Ballantyne+06}; \citealt{Hasinger+08}; \citealt{DellaCeca+08}: \citealt{Treister+09}; \citealt{Iwasawa+12}); on the other hand, the space density of the obscured ($\log [N_{\rm H} / \rm cm^{-2}] > 22 $) AGN population seems to be consistent with a no evolution scenario, at least for low-luminosity ($L_{\rm X} < 10^{44}$ erg s$^{-1}$) X-ray AGN (\citealt{Ueda+03}; \citealt{Akylas+06}; \citealt{Gilli+10}; \citealt{Vito+13}).

AGN are ubiquitous X-ray emitters (\citealt{Elvis+78}) and X-ray surveys are only weakly affected by the X-ray host galaxy light. Nevertheless, recent works (\citealt{Treister+12}; \citealt{Schawinski+12}; \citealt{Laird+10}) claim that the host galaxies of high-redshift ($z > $1--2) AGN could have experienced intense bursts of star formation (e.g. Sub-Millimeter Galaxies, SMGs and/or Ultra-Luminous Infrared galaxies, ULIRGs), likely hiding heavily obscured AGN with intrinsically high bolometric luminosities, but enshrouded in dense, cold, optically thick gas clouds (see \citealt{Alexander+05}). 

Infrared (IR) observations are only mildly affected by obscuration and can detect warm dust signatures for both unobscured and obscured AGN, giving rise to a complementary perspective to the X-ray one. This is the reason why IR observations represent a crucial tool to investigate and constrain the obscured growth across cosmic time. Recent \textit{Herschel}-based studies are shedding light on the mid-to-far infrared properties of star-forming galaxies (\citealt{Hatziminaoglou+10}; \citealt{Rowan-Robinson+10}; \citealt{Hwang+10}; \citealt{Elbaz+11}; \citealt{Sajina+12}; \citealt{Smith+12}; \citealt{Symeonidis+13}; \citealt{Magdis+13}), both in presence and in absence of a significant AGN emission component. The importance of IR observations in chasing obscured and unobscured AGN has also been proved through mid-infrared (MIR) photometry, which allows us to develop interesting diagnostics to detect AGN features (\citealt{Lacy+04}; \citealt{Stern+05}; \citealt{Pope+08}; \citealt{Donley+12}; \citealt{Juneau+13}; 
\citealt{Lacy+13}). However, these techniques have still to be tested on high redshift ($z > 1$) sizeable samples (\citealt{Juneau+11}; \citealt{Trump+13}) and are sensitive to comparatively large AGN-to-host flux ratios in the MIR domain (\citealt{Barmby+06}; \citealt{Brusa+10}; \citealt{Donley+12}). Hence, possible signatures of nuclear accretion for both intrinsically low-luminosity and heavily absorbed AGN might be diluted by the host galaxy light. These complications highlight the need for a complete assessment of the amount of star formation in AGN host galaxies.

The main purpose of this paper is to provide an estimate of the BHAD from an IR survey through a multi-wavelength analysis of galaxy SEDs. The \textit{Herschel} space observatory (\citealt{Pilbratt+10}) gives an unique opportunity to constrain the far-infrared (FIR) peak of star-forming galaxy SEDs, which can be used as a proxy for their star formation rate (SFR), down to unprecedented flux densities ($\simeq $ few mJy), reaching to $z > 3$. 

The FIR peak is mainly due to optical/UV stellar light reprocessed by dust. However, an additional contribution can arise from warmer nuclear dust heated by material accreting onto the SMBH rather than stellar processes. \textit{Spitzer} data, together with \textit{Herschel} capabilities, are paramount to reach a complete assessment of the dust content at different temperatures stored across the IR domain. \textit{Herschel} does play a key role in investigating the star formation/AGN connection down to intrinsically low luminosity sources without the need to extrapolate previous stellar+dust emission templates to rest-frame FIR wavelengths. This is the reason why we exploit observations carried out within the \textit{\rm PACS} Evolutionary Probe (PEP\footnote{\url{http://www2011.mpe.mpg.de/ir/Research/PEP/} }, \citealt{Lutz+11}) and the \textit{Herschel} Multi-tiered Extragalactic Survey (HerMES, \citealt{Oliver+12}) in two different regions: the \textit{Great Observatories Origins Deep Survey-South} (GOODS-
S, \citealt{Elbaz+11}), the deepest, pencil beam, survey observed by \textit{Herschel}, and the \textit{Cosmic Evolution Survey} (COSMOS, \citealt{Scoville+07}), one of the widest among the PEP fields. Such a combination allows us to sample different luminosity regimes and to span a wide range of redshifts, while preserving a sizeable number of sources.

The paper is structured as follows. The sources in the \textit{Herschel}-selected sample are described in \S~2, together with a discussion on their counterparts in other wavelengths and their redshift distribution; in \S~3 we describe our approach to decouple the AGN contribution from that related to SF within each global SED and how we evaluate the relative incidence of the nuclear content. In \S~4 we derive the AGN bolometric luminosity function up to $z\sim$3, while in \S~5 we present the BHAD evolution over cosmic time, discussing our results and comparing them with previous estimates from the literature. In \S~6 we present our main conclusions.

Throughout this paper, we adopt a \citet{Chabrier03} initial mass function (IMF) and we assume a flat cosmology with $\Omega_m = 0.27$, $\Omega_\Lambda = 0.73$, and $H_0 = 71$ km\,s$^{-1}$\,Mpc$^{-1}$.

\section{Sample Description} \label{sample}

In this work we exploit large photometric data-sets, from the UV to the sub-mm, both in GOODS-S and in COSMOS fields. We take advantage of PACS (\citealt{Poglitsch+10}) photometry in the FIR (100 and 160$~\mu$m; 70$~\mu$m data are available for the GOODS-S field only). As a reference sample we use the PEP 160$~\mu$m blind catalogue (internal version 1.2 in GOODS-S and version 2.2 in COSMOS). The PEP sample has sources selected above 3$\sigma$ detection at 160$~\mu$m, corresponding to nominal flux density limits of 2.4 and 10.2 mJy in the GOODS-S and COSMOS fields, respectively. Globally we find 782 and 5105 sources at 160$~\mu$m, in the GOODS-S and the COSMOS fields, respectively. The available photometry has been extended up to sub-mm wavelengths (250, 350 and 500$~\mu$m) through the \textit{Spectral and Photometric Imaging Receiver} (SPIRE, \citealt{Griffin+10}) data of the HerMES survey (\citealt{Oliver+12}). 

Here we briefly outline the multi-band identification procedure, as well as the overall redshift distribution, referring to \citeauthor{Berta+10} (\citeyear{Berta+10}, \citeyear{Berta+11}) and \citet{Lutz+11} for further details concerning data reduction, completeness and evaluation of spurious sources and photometric redshift uncertainties.

\subsection{Multi-wavelength identification} \label{sample_multil}

PEP-selected sources have been cross-matched with their respective counterparts using a maximum likelihood technique (\citealt{Sutherland+92}; \citealt{Ciliegi+01}). Starting from each PACS 160$~\mu $m detection, we checked for counterparts at shorter wavelengths through a cross-match with detections at 100$~\mu $m, 70$~\mu $m (in the GOODS-S sample only) and down to 24$~\mu$m, still using a maximum likelihood algorithm.

In GOODS-S we have matched the PACS area in common with the inner region covered by multi-wavelength observations in UV, optical and near-IR. The resulting common area (196 arcmin$^2$) is around 65 per cent of the parent sky area covered by PACS. The number of PACS sources within the inner region is 494, which is around 63 per cent of the parent PACS population. This suggests that the surface density of PACS sources remains almost unchanged. In this inner GOODS-S area, the \textit{Spitzer}-MIPS 24$~\mu$m selected catalogue (\citealt{Magnelli+09}) is linked to \textit{Spitzer}-IRAC 3.6$~\mu $m positions through prior extraction. MIPS 24$~\mu $m detections act as a connecting link between the FIR and optical regimes, since they have been cross-correlated with the \textit{MUlti-wavelength Southern Infrared Catalog} (MUSIC \footnote{The MUSIC catalogue can be retrieved at \url{http://lbc.oa-roma.inaf.it/goods/goods.php}}, \citealt{Grazian+06}; \citealt{Santini+09}). 

The MUSIC spectral coverage ranges from the far-UV ($\sim$3500 \AA) up to the mid-IR (\textit{Spitzer}-IRAC at 8$~\mu$m). As mentioned before, to gather as much information as possible for each source, we consider the \textit{Herschel}-selected sample within the MUSIC area (196 arcmin$^2$). 99 per cent of these sources (494 at 160$~\mu$m) have a counterpart in the 24$~\mu$m catalogue and in the optical (MUSIC catalogue), above their corresponding flux density limits. Such high percentages of identified counterparts suggests that 24$~\mu $m and optical observations in the GOODS-South field are deeper than the \textit{Herschel}-PACS ones, with respect to an average MIR-to-FIR galaxy SED.

We have also considered HerMES observations (\citealt{Oliver+12}), which entirely cover the PEP area with SPIRE maps at 250, 350 and 500$~\mu $m. Also the SPIRE counterparts of our PACS-selected sample have been restricted to the common MUSIC area. In HerMES, SPIRE flux densities are taken by following the prior source extraction presented by \citet{Roseboom+11}, based on \textit{Spitzer}-MIPS 24$~\mu$m positions as a prior, and cross-matched to the blind PACS catalogue. In the GOODS-S area, we have also considered \textit{Spitzer}-IRS 16$~\mu$m detections (\citealt{Teplitz+11}), which have been cross-correlated to the available 24$~\mu$m positions through a nearest neighbour match.

In the COSMOS field, the sky area explored by \textit{Herschel} and observations at shorter wavelengths is around 2 deg$^2$. PACS 160$~\mu$m sources have been associated with the \textit{Spitzer}-MIPS 24$~\mu$m ones from \citet{LeFloc'h+09} and with the IRAC-selected catalogue from \citet{Ilbert+10}, which is already matched to UV, optical and near-IR photometry. By removing PEP sources within flagged areas of the optical-to-NIR catalogues, we are left with 4118 sources at 160$~\mu$m identified at shorter wavelengths. Around 96.5 per cent of them have a MIPS 24$~\mu $m detection in the common area. The fraction of PACS-selected sources having at least one SPIRE counterpart ($>3\sigma$) reaches 84 per cent in the GOODS-S field (within the MUSIC area) and 86 per cent in the COSMOS field. 

In both fields, we checked some properties of PACS sources without SPIRE counterpart and we ensured that the integrated IR (8--1000$~\mu $m) luminosities\footnote{Calculated through SED-fitting. See Section \ref{fitting} for details.} and redshifts were consistent within 1$\sigma$ uncertainty with those of PACS sources having SPIRE counterpart.

Overall, we collect photometry at a maximum of 22 (GOODS-S) and 20 (COSMOS) different wavebands, from UV to FIR wavelengths. The numbers of PEP-selected objects fulfilling the optical identification criterion are 494 and 4118, in the GOODS-S and COSMOS, respectively.

\subsection{Redshift distribution} \label{sample_z}

\begin{figure}
\begin{center}
    \includegraphics[width=\linewidth]{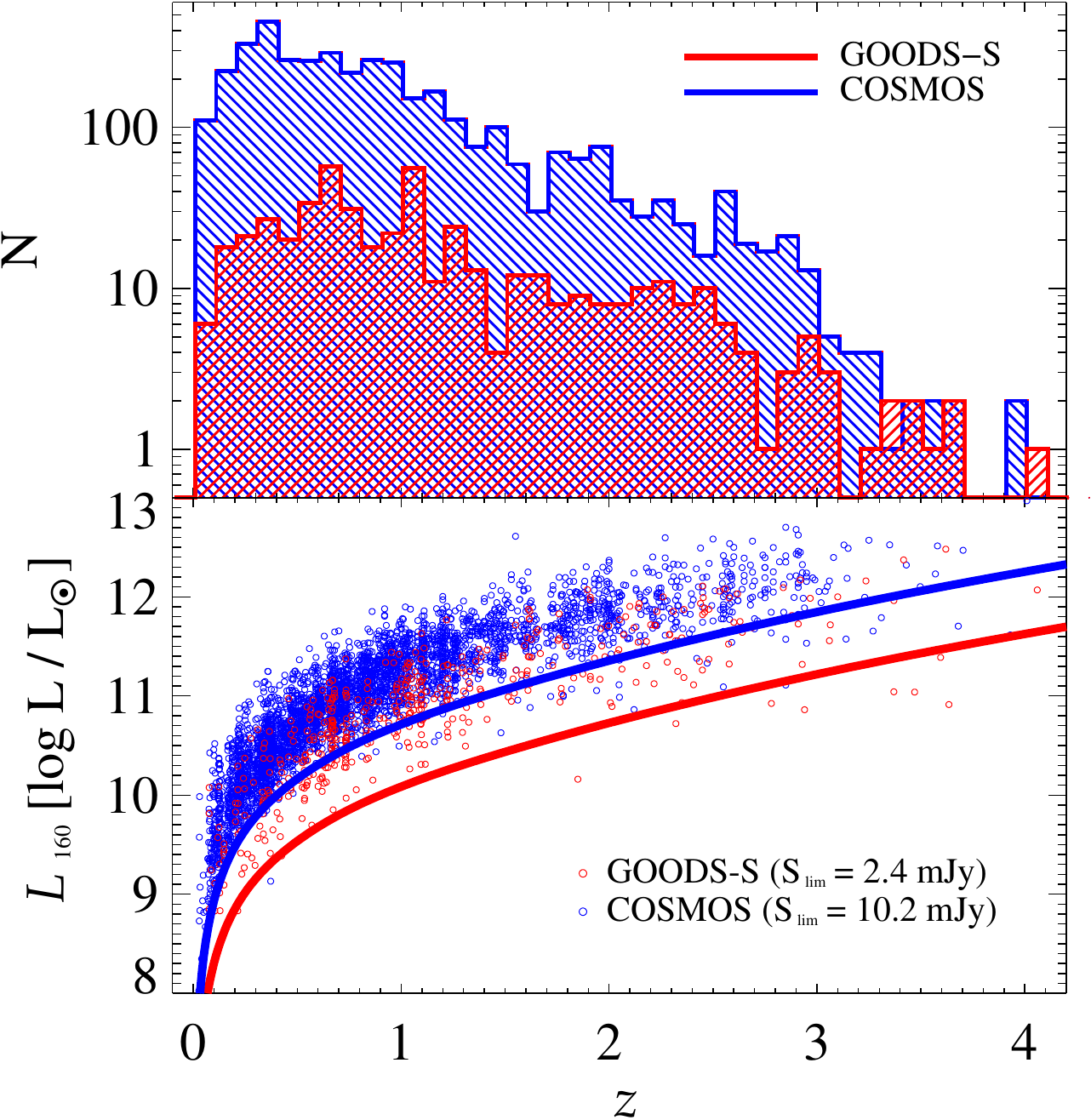}
\end{center}
 \caption{\textit{Top panel}: redshift distribution of the 160$~\mu$m PEP sources identified in optical bands, in GOODS-S (\textit{red dashed}) and COSMOS (\textit{blue dashed}). Note that the scale of the \textit{y}-axis is logarithmic. 
 \textit{Bottom panel}: infrared luminosity limit at 160$~\mu$m (rest-frame) as a function of redshift. Open circles represent the luminosity limits for GOODS-S (\textit{red}) and COSMOS (\textit{blue}) sources. The solid curves show the minimum luminosity (corresponding to the limiting flux) versus redshift for the two fields (GOODS-S in red, COSMOS in blue), obtained for the empirical SED of the source IRAS 20551--4250. }
   \label{fig:z_dist}
\end{figure}

In GOODS-S we rely on a large spectroscopic database,\footnote{Publicly available at: \url{http://www.eso.org/sci/activities/garching/projects/goods/MasterSpectroscopy.html} } including more than 3000 sources (\citealt{Cristiani+00}; \citealt{Croom+01}; \citealt{Bunker+03}; \citealt{Dickinson+04}; \citealt{Stanway+04b,Stanway+04a}; \citealt{Strolger+04}; \citealt{Szokoly+04}; \citealt{vanderWel+04}; \citealt{Doherty+05}; \citealt{LeFevre+05}; \citealt{Mignoli+05}; \citealt{Vanzella+08}; \citealt{Popesso+09}; \citealt{Santini+09}; \citealt{Balestra+10}; \citealt{Cooper+12}; \citealt{Kurk+13}). We used this to enrich the existing spectroscopic database of the MUSIC catalogue (\citealt{Grazian+06}; \citealt{Santini+09}) replacing MUSIC photometric redshifts with the up-to-date spectroscopic ones in the case of ``secure'' redshift determination (according to the scale adopted by \citealt{Balestra+10}). In summary, the GOODS-S PEP sample includes 494 sources with 100 per cent redshift completeness (71 per cent 
spectroscopic and 29 per cent photometric).

A rich redshift data-set is also available for COSMOS, including both photometric (\citealt{Ilbert+10}) and spectroscopic (\citealt{Lilly+09}; \citealt{Trump+09}) redshifts. 

Recently, \citet{Berta+11} extended the previous photometric redshift data-set for most of the remaining PEP sample, obtaining an overall redshift completeness as high as 93 per cent (about 40 per cent spectroscopic). By using the latter data-set, we gather a number of 3849 PEP sources with a redshift estimate.

The photometric redshift uncertainties have been evaluated in both fields from previous works. In GOODS-S, \citet{Grazian+06} found an average absolute scatter $\left \langle |\Delta z/(1 + z)| \right \rangle =$ 0.045 over the whole redshift range 0$<$\textit{z}$<$6. In COSMOS, \citet{Ilbert+10} evaluated the scatter on the photo-\textit{z}s to range from 0.008 (for $I < 22.5$) to 0.053 (for $24 < I < 25$), while the more up-to-date photometric redshifts provided by \citet{Berta+11} have a median absolute deviation (MAD\footnote{{MAD$(x)$ = median($|(x)$ - median$(x)|$)}, where \textit{x} represents the difference between spectroscopic and photometric redshift.} ) of around 0.01. The expected fraction of outliers, defined as objects having $\langle |\Delta z/(1 + z)| \rangle \geq $ 0.2, is as high as $\sim$2 per cent.

The final sample of GOODS-S (GS) and COSMOS (C) sources with redshifts includes 4343 objects (494 for GS and 3849 for C), which are selected at 160$~\mu$m. The global redshift distributions of PACS-160$~\mu$m selected sources are presented in Fig. \ref{fig:z_dist} (top) for both the PEP fields. As expected, because of the higher flux limit, the COSMOS redshift distribution has a lower median redshift value ($\left \langle z \right \rangle \sim 0.7$) than the GOODS-S ($\left \langle z \right \rangle \sim 1$), although it also extends to high redshift ($z>3$). To highlight the range of infrared luminosities sampled at different redshifts, we show in Fig. \ref{fig:z_dist} (bottom) the minimum 160$~\mu$m luminosity versus redshift of the (rest-frame) 160$~\mu$m luminosity. Red and blue circles represent GOODS-S and COSMOS sources, respectively. The solid curves represent the minimum luminosity (corresponding to the limiting flux) versus redshift for the two fields (GOODS-S in red, COSMOS in blue). This curve is 
purely illustrative and it is taken from the SED template of \textit{IRAS} 20551--4250, which is a local star-forming galaxy hosting an AGN (see \citealt{Polletta+07}).

\section{Fitting broad-band SED$_{\textbf{s}}$}  \label{fitting}

In this section we present the SED-fitting analysis performed for the 4343 PACS sources with optical-to-MIR counterparts and either spectroscopic or photometric redshift. Each observed SED has been fitted making use of the {\sc magphys} code\footnote{{\sc magphys} can be retrieved at \url{http://www.iap.fr/magphys/magphys/MAGPHYS.html} } (\citealt{DaCunha+08}) and a modified version of this code presented by \citet{Berta+13}.

\begin{figure*}
\begin{center}
    \includegraphics[width=180mm,keepaspectratio]{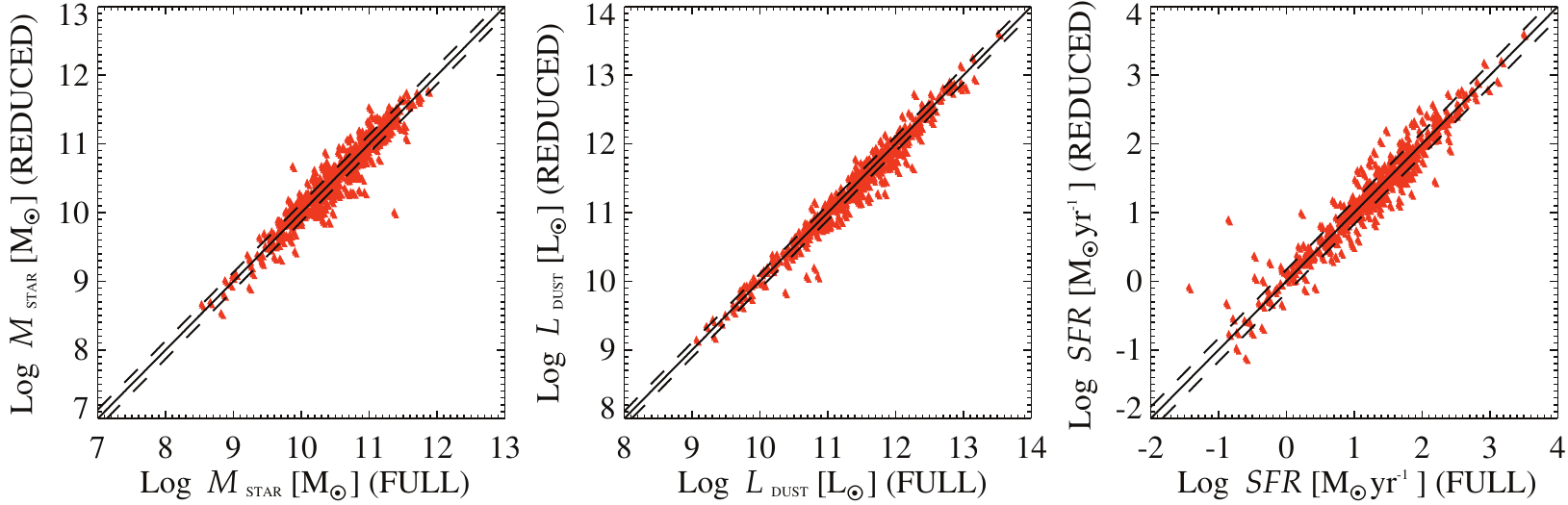}
\end{center}
 \caption{Comparison between the values obtained with the full (x-\textit{axes}) and reduced (y-\textit{axes}) {\sc magphys} library of templates for three different integrated parameters: stellar mass; infrared (8--1000$~\mu$m) luminosity due to star formation; and star formation rate. The full grid uses all the available templates of galaxy SEDs encoded in {\sc magphys}, whereas the reduced grid accounts for a small fraction of them resulting from a random selection (see text for details). Red triangles are the best-fit values of each parameter as taken from the GOODS-S sample. The black solid line is the bisector, while the dashed lines represent the $\pm$ 1$\sigma$ confidence regions. The resulting 1$\sigma$ uncertainties related to these parameters are 0.13, 0.11 and 0.17 dex, respectively.}
   \label{fig:magphys_grid}
\end{figure*}

\subsection{SED-fitting with {\sc \textbf{magphys} } } \label{sed_noagn}

{\sc magphys} is a public code which uses physically motivated templates to reproduce the observed SEDs from the ultraviolet to sub-mm wavelengths. It has been developed especially to fit star-forming galaxy SEDs, so that it is well suited for our purposes (see e.g. \citealt{Smith+12} for an extensive application of {\sc magphys} on \textit{Herschel} data). Indeed, the basic recipe that this code relies on is a self-consistency between the energy originating from stars, which is partially absorbed by dust, and the infrared light due to re-emission by the dust itself. Briefly, the stellar and dust component are computed as follows (see \citealt{DaCunha+08} for details).
\begin{itemize}
 \item The \citet{Bruzual+03} library of templates is used to compute the integrated light produced by stars in galaxies. These models predict the spectral evolution of stellar populations at $\lambda \leq $ 10$~\mu $m and ages between $10^{5}$ and $10^{10}$ yr, starting from a \citet{Chabrier03} IMF and combining parameters representative of different metallicities, star formation histories and dust contents.
\item Each stellar template is modified by the effect of dust extinction, which is parametrised as the cumulative attenuation produced by interstellar medium (ISM) and dusty molecular clouds, according to the angle-averaged model of \citet{Charlot+00}. 
\item The global dust emission is ascribed to three main constituents of interstellar dust: polycyclic aromatic hydrocarbons (PAHs, in the range 3$-$12$ \mu$m); mid-IR continuum from hot  (130$-$250 K) dust grains; and thermal emission, both from warm (30$-$60 K) and cold (15$-$25 K) dust grains in thermal equilibrium, according to \citet{DaCunha+08} prescriptions.
\end{itemize}

The analysis of each single observed spectral energy distribution with {\sc magphys} starts by building up a large library of templates at the same redshift of the source. Given the set of parameters considered as well as the wide range of values adopted by each of them, in the end 50,000 stellar and 50,000 independent dust emission models are computed for each source. Only about 30 per cent of the IR templates can be combined with each stellar model, because of the underlying assumption that all the absorbed UV and optical stellar radiation has to be accounted for in the integrated infrared emission. As a result of such selection, tipically 10$^9$ different self-consistent (i.e. stellar+dust) model combinations are created to reproduce each observed SED from the UV to the FIR. Then, each model combination is compared to the observed SED and the stellar+dust model normalization is rigidly scaled to fit the data as well as possible. Moreover, a Bayesian approach is implemented to build a marginalised 
likelihood distribution for each parameter, which is traced by the distribution of values of such parameter across the whole library of models. This statistical analysis accounts for the degeneracy level in the parameter space, which occurs when a relatively small set of data points is compared with such a large range of models. The resulting probability distribution function (PDF) allows the user to obtain reliable confidence ranges for parameter estimates.

\subsection{Fitting with {\sc \textbf{magphys + agn} } } \label{sed_agn}

{\sc magphys} does not consider a potential AGN contribution, but only star formation processes are involved in building the SEDs. Since we are interested in probing the possible AGN contribution to the observed SEDs, we make use of a modified version of the code (\citealt{Berta+13}), which accounts also for a possible nuclear heating source.

The code by \citet{Berta+13} includes the AGN component and differs from the original {\sc magphys} in the model normalization approach. The stellar+dust normalization is not scaled to fit the data, but is allowed to float within a range of normalization values (the same range adopted by the original version of {\sc magphys}), to allow for the additional AGN normalization when fitting the observed SED. This approach results in a three component simultaneous fit, using an overall $\chi^2$ minimization. 

The AGN library of templates (\citealt{Feltre+12}; see also \citealt{Fritz+06}) includes around 2400 models, each one analysed for ten different lines of sight $\theta$, uniformly distributed in $\theta$ from edge-on to face-on. This set of templates is widely used in the literature, since it is physically motivated and suitable in reproducing a wide variety of AGN features, such as hot circumnuclear dust in mid-IR dominated SEDs (e.g. Mrk3, see \citealt{Mullaney+11}) and the silicate features at 9.7 and 18$~\mu$m (see \citealt{Feltre+12}).

Each nuclear template includes both the contribution from the central engine (i.e. accretion disc) and the reprocessed emission from a smooth sorrounding dusty torus.

The central heating source is modeled with a broken power law $\lambda \rm L_{\lambda} \propto {\lambda}^\alpha$, with slopes $\alpha_1$ = 1.2, $\alpha_2$ = 0, $\alpha_3$ = $-$1, in the wavelength ranges $\lambda_1$ = 0.001--0.03$~\mu$m (\citealt{Hubeny+00}), $\lambda_2$ = 0.03--0.125$~\mu$m (\citealt{Zheng+97}), and $\lambda_3$ = 0.125--20$~\mu$m (\citealt{Hatziminaoglou+08}), respectively. At longer wavelengths the emission scales as a Planck spectrum in the Rayleigh-Jeans regime. The implementation of the circumnuclear dust component relies on the numerical solution of the radiative transfer equation for a smooth dusty structure.

Each torus model is univocally identified by a set of six different parameters: the outer to inner radius ratio $R_{\rm out}/ R_{\rm inn}$; the opening angle $\Theta$; the equatorial optical depth at 9.7$~\mu$m $\tau_{9.7}$; the radial and height slopes of the density profile ($\beta$, $\gamma$); and the viewing angle $\theta$ of the line of sight.

During the AGN fitting procedure, 100 different AGN templates are randomly selected for each source and for each stellar+dust realization. In this way it is possible to optimise the random sampling by exploring a wider range of the AGN parameter space for each source.

By solving the radiative transfer equation for a smooth dusty structure irradiated by the accretion disc, a bolometric correction (BC) value is available for each AGN template. Such information allows us to calculate the AGN bolometric luminosity ($L_{\rm bol, AGN}$) directly from the infrared (1--1000$~\mu$m rest frame) luminosity of the best-fit AGN model, according to the following expression:
\begin{equation}
k_{\rm bol} = \frac{L_{\rm bol, AGN}}{L^{\rm agn}_{1-1000}}
\end{equation}
Our $L_{\rm bol, AGN}$ estimates are derived without double counting the dust-reprocessed contribution arising from the accretion disc, since the radiative transfer recipe performs a self-consistent re-distribution of the input energy at all wavelengths, according to the adopted set of templates.

\subsubsection{Reduced library of templates}    \label{reduced_grids}

The set of possible model combinations included in the original version of {\sc magphys} is very large (around $10^9 $ possible configurations, see \S~{\ref{sed_noagn}}), as well as the set of physical parameters that the stellar+dust emission models rely on. The range of values spanned by each parameter is physically reasonable, but the computational time taken by the code strongly increases when the AGN component is added to the SED-fitting procedure. For these reasons we take advantage of a reduced grid of templates, as described by \citet{Berta+13}.

While the full grid of {\sc {magphys}} templates includes 50000$\times$50000 model configurations, the reduced grid has been built up through a random selection of 1000 stellar templates for each source and 1000 dust-emission templates for each stellar template. This selection technique permits to save a factor of 2500 in time (2 hours per object, instead of 5000), giving a total of $\sim$3$\times 10^5$ self-consistent templates for each source (after the 30 per cent cut required by the energy balance argument). Such selection uniformly spans the parameter space covered by the original library, without introducing any significant bias in the adopted grid. In Fig. {\ref{fig:magphys_grid}} we compare the values of some integrated physical 
quantities (star formation rate, infrared luminosity, etc.) obtained using the full and the reduced grid of {\sc magphys}. The plots show that mostly the integrated physical parameters are consistent among the full and reduced grids. 

As described in \S~{\ref{sed_agn}}, each AGN template is defined by the combination of six physical parameters. We limit the torus library of templates by restricting the multi-dimensional volume of the parameter space. In particular, we removed the largest optical depth value at 9.7$~\mu$m (see \citealt{Pier&Krolik92}) and the most extended geometries ($R_{\rm out}/ R_{\rm inn} > $100), as no evidence for such large structures has yet been found (\citealt{Jaffe+04}; \citealt{Tristram+07}, \citeyear{Tristram+09}). 

After this cut, the torus grid includes 504 templates, each one computed at 10 different lines of sight. As mentioned before, a random selection of 100 / 5040 AGN configurations is performed for each source and stellar+dust model combination. This reduced library spans several values of the ratio $R_{\rm out}/ R_{\rm inn}$ (10, 30, 60 and 100), the opening angle $\Theta$ (40$^{\circ}$, 100$^{\circ}$ and 140$^{\circ}$), the equatorial optical depth $\tau_{9.7}$ (0.1, 0.3, 0.6, 1, 2, 3 and 6) and the slopes of the density profile (--1, --0.5 and 0 for $\beta$; 0 and 6 for $\gamma$). The resulting library turns out to be similar to the one used by \citet{Pozzi+12} and \citealt{Hatziminaoglou+08}, \citeyear{Hatziminaoglou+09}), which we refer to for a detailed description of the degeneracy level drawn by torus parameters.

\begin{figure}
\begin{center}
    \includegraphics[width=\linewidth]{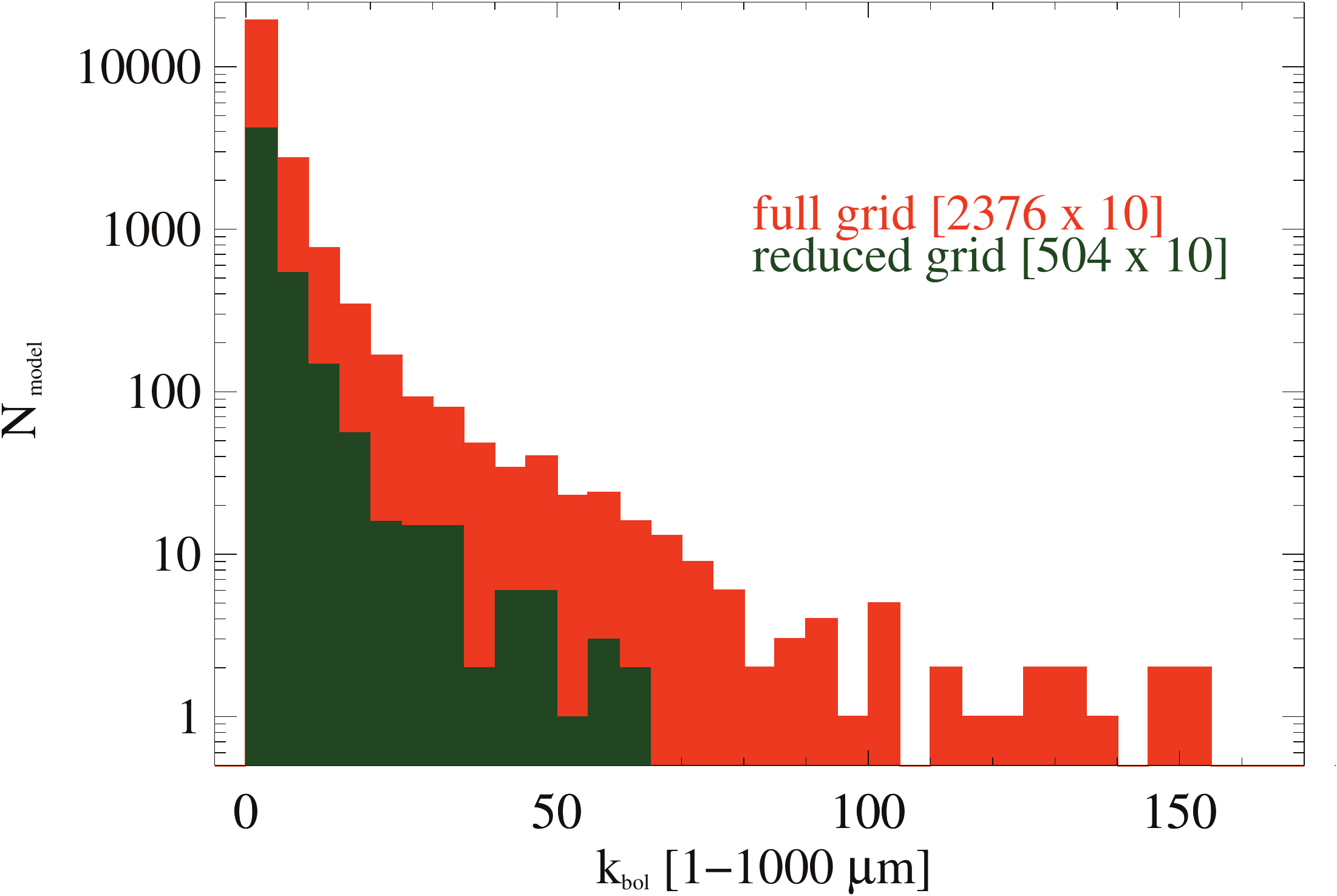}
\end{center}
 \caption{Bolometric correction (1--1000$~\mu$m, rest-frame) distributions to turn the infrared AGN luminosity into bolometric luminosity. The red area traces the full grid of the library by \citet{Feltre+12} and \citet{Fritz+06}. The green area represents the reduced library that is adopted in this paper. }
   \label{fig:kbol_model}
\end{figure}

Since the only crucial parameter in the following analysis is the AGN bolometric correction ($k_{\rm bol}$), used to convert the total infrared AGN emission into AGN bolometric luminosity, we show in Fig. \ref{fig:kbol_model} how much our restrictions modify the shape of the distribution of the bolometric corrections. In particular, the reduced grid of AGN templates reaches $k_{\rm bol}$ values around 60, while the full grid samples up to $k_{\rm bol} \sim 150$. The largest values are mostly representative of the most obscured AGN ($\tau_{9.7}$=10). Despite the existing cut in $k_{\rm bol}$, we point out that the the missed values (i.e. $k_{\rm bol} > 60$) constitute a negligible number of AGN templates with respect to the full grid ($\sim$ 0.3 per cent). 

In addition, we checked the stability of our $k_{\rm bol}$ estimates with increasing number of galaxy templates. We have run the {\sc magphys+agn} SED-fitting on a subsample of 50 \textit{Herschel} sources randomly selected from the GOODS-South field. We compared the $k_{\rm bol}$ estimates taken from the run with 1000$\times$1000  {\sc magphys} templates with those obtained from the run with 5000$\times$5000 templates. We found no systematics between the two SED-fitting analyses, with a 1$\sigma$ scatter equal to 0.07 dex. Though this test does not account for the full grid of available galaxy templates, we can at least say that the AGN bolometric correction does not change significantly with increasing number of galaxy templates.

\subsubsection{AGN Bolometric Luminosity and Bolometric corrections} \label{bol_info}
 
For every observed SED, a PDF is created for each parameter to trace its confidence range. As mentioned before, among the physical parameters returned by the code, we have focused on the AGN bolometric luminosity only (computed by accounting for the rest-frame 1-1000$~\mu $m AGN emission and adopting its related bolometric correction).

We stress that each $L_{\rm bol, AGN}$ value is not properly a bolometric luminosity, since the input (i.e. accretion) energy of the central engine is limited to the range 10$^{-3}$ -- 10$^3\mu $m (i.e. 10$^{-1}$ -- 10$^{-7}$ keV). This means that the X-ray emission is supposed to be negligible in terms of input energy. Such an assumption relies on the fact that large $k_{\rm bol}$ values have been found in hard X-rays, of the order of 20$-$30 (\citealt{Risaliti+04}; \citealt{Marconi+04}; \citealt{Hopkins+07}; \citealt{Pozzi+07}; \citealt{Vasudevan+09}; \citealt{Lusso+12}), increasing as a function of $L_{\rm bol, AGN}$.\footnote{The terms ``accretion'' and ``bolometric'' AGN luminosity adopted throughout the paper are assumed to have the same physical meaning.}

\begin{figure}
\begin{center}
      \includegraphics[width=\linewidth]{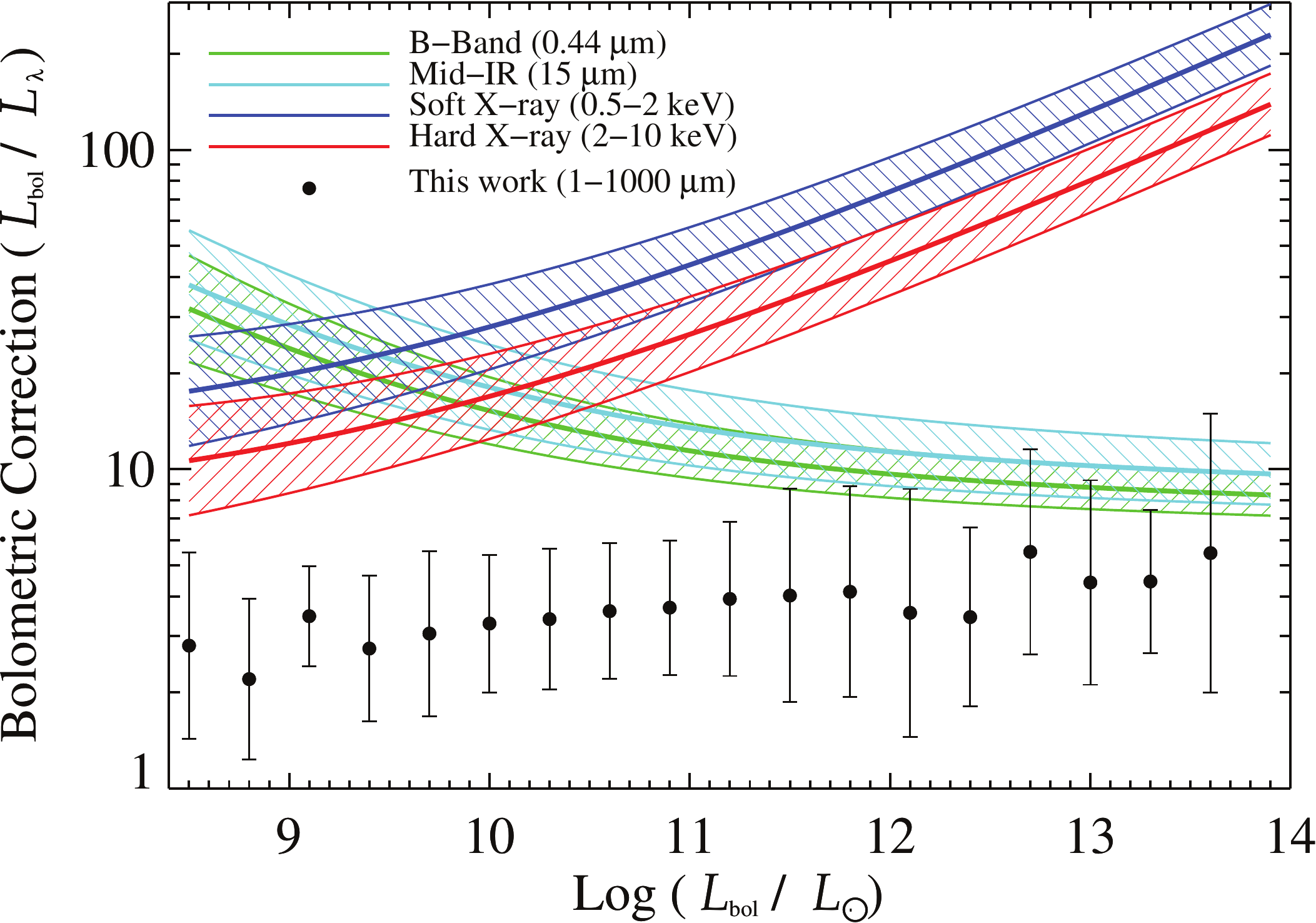}
\end{center}
 \caption{Distributions of AGN bolometric corrections. The regions traced by solid lines represent various trends of $k_{\rm bol}$ computed in different bands by \citet{Hopkins+07}. Black circles (with $\pm 1\sigma$ uncertainties) come from the sample analysed in this paper, only including sources with relevant signatures of nuclear activity (see \S~\ref{f_test} for details). }
   \label{fig:kbol}
\end{figure}

In Fig. \ref{fig:kbol} we show a comparison between distributions of bolometric corrections computed in various bands. On the one hand, \citet{Hopkins+07} derived their $k_{\rm bol}$ values by assuming different amounts of obscuration applied to an average, unobscured, type-1 AGN SED (\citealt{Richards+06}). They found $k_{\rm bol}$ to depend on the input AGN luminosity, consistent with other works (e.g. \citealt{Marconi+04}; \citealt{Lusso+12}). On the other hand, we derive the distribution of $k_{\rm bol}$ by integrating the best-fit AGN model of each source (if it fulfills the criterion imposed by the Fisher test, as explained in \S~\ref{f_test}) over the whole IR domain (1--1000$~\mu$m, rest-frame). Each black circle represents the median $k_{\rm bol}$ value on different luminosity bins ($=0.3$ dex wide). Error bars correspond to $\pm 1\sigma$ uncertainties and are centered on the median value of the respective bin. We find that $k_{\rm bol}$ does not change significantly with increasing AGN bolometric 
luminosity $L_{\rm bol, AGN}$. This result could suggest that the fraction of the input energy coming from the central engine which is absorbed by the obscuring structure and re-emitted in the IR does not change significantly as a function of $L_{\rm bol, AGN}$.

The average bolometric corrections derived from the IR are much lower than those derived from X-ray measurements (see Fig. \ref{fig:kbol}). Indeed, the distribution of 1--1000$~\mu$m bolometric corrections has a median value of $\sim$3.8 over the whole range of $L_{\rm bol, AGN}$ covered by our sample, ranging on average from 2 to 5.

\subsection{Testing the AGN incidence}   \label{f_test}

\begin{figure*}
\begin{center}
    \includegraphics[width=180mm,keepaspectratio]{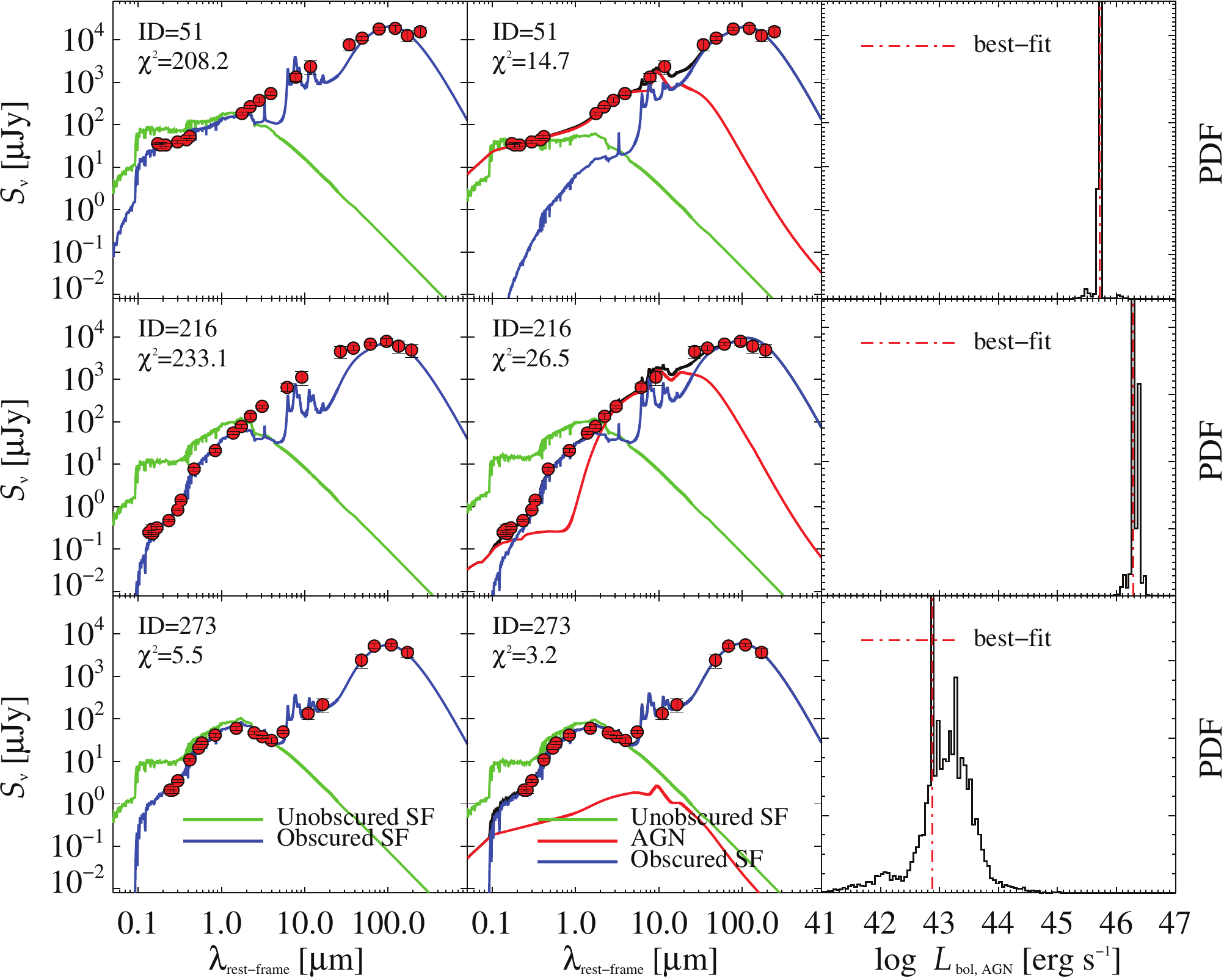}
\end{center}
 \caption{Three examples of SED decomposition performed without AGN (\textit{left panels}) and with AGN (\textit{central panels}). The corresponding PDF of the AGN bolometric luminosity is shown in the right panels. The red circles represent the observed data points. The green line shows the integrated unextincted emission originating from stars. The blue solid line represents the star formation contribution for dust absorption, partially redistributed across the MIR/FIR range in a self-consistent way. The red line reproduces the AGN contribution and incorporates both the accretion disc and the torus emission, according to the adopted library of templates. The sources shown in the upper and middle panels are classified as AGN at $\geq$99 per cent confidence, while one example of source will less significant AGN contribution is shown in the bottom panel.}
   \label{fig:fit_magphys}
\end{figure*}

In order to evaluate the impact of an additional AGN component on a ``pure'' star-forming galaxy SED, for each source we run both {\sc magphys} and {\sc magphys+agn}. As mentioned in \S~{\ref{reduced_grids}}, a reduced grid has been adopted in both fitting procedures. First, we run the code version including the AGN and during the random selection of stellar and dust templates their template identification numbers are registered. Once the SED-fitting with {\sc{magphys+agn}} is accomplished, we run the {\sc{magphys}} code for each source by using the same set of stellar+dust templates previously extracted for the same source. This implementation allows us to test if the fit provided by similar star formation templates improves when accounting for an additional nuclear emission component.

In Fig. {\ref{fig:fit_magphys}} three different examples of best-fits are shown as a result of this SED decomposition. The integrated emission has been decomposed into three emission contributions, according to the recipe previously described. The expected stellar light (green line) is partially absorbed by a mixture of dust grains and re-distributed in the MIR/FIR domain (blue line). The AGN emission is the red curve. The respective $\chi^2$ values are then used to estimate the significance of the improvement of the fit occurring when a nuclear component is added.

Around 98 per cent of the PEP sample fitted with the {\sc magphys+agn} code includes an AGN component in the best-fit model. However, the addition of the AGN component does not always significantly improve the best-fit. For instance, the last source shown in Fig. {\ref{fig:fit_magphys}} turns out to exhibit negligible signatures of AGN activity over the entire wavelength range. To pinpoint the significance of the AGN component on the global SED we apply a Fisher test (\textit{F}-test) to the two SED-fitting runs, as presented by \citet{Bevington+03}:

\begin{equation}
F_{\rm test} = \frac{\chi^{2}_{\rm no~ agn} - \chi^{2}_{\rm agn}}{\chi^{2}_{\nu,~ \rm agn}} ~\geq ~ F_{\rm threshold} (\rm CL)
   \label{eq:f_test}
\end{equation}
with $\chi^{2}_{\rm no~ agn}$ and $\chi^{2}_{\rm agn}$ being the $\chi^2$ values of the best-fit without and with the AGN, respectively, whereas ${\chi^{2}_{\nu,~ \rm agn}}$ is the reduced $\chi^{2}$ value referred to the best-fit with an AGN component. If this ratio exceeds a given threshold, then the source of interest is included in the following analysis, otherwise it is ruled out and a null AGN contribution is assumed for its specific SED. The threshold value depends on the number of degrees of freedom as well as on the required confidence level (CL).

If carrying out this analysis at a 99 per cent confidence level, in the GOODS-S field we find that the AGN detection rate is 48 per cent, while in COSMOS it is around 36 per cent. By evaluating the binomial confidence limits following \citet{Gehrels86} and \citet{Gerke+07}, such difference turns out to be significant at the $>3\sigma$ level. To make a more coherent comparison between these percentages, we checked the incidence of some possible selection effects. We find that the AGN detection rate depends on both the observed 160$~\mu $m flux density and the presence of the 16$~\mu $m detection, which is missing in COSMOS. As a sanity check, we run {\sc magphys} and {\sc magphys+agn} on the entire GOODS-S sample without using the 16$~\mu $m data during the SED-fitting. To compare the resulting AGN detection rate with that obtained in COSMOS, we selected only sources above the COSMOS 160$~\mu $m flux density limit ($S_{160} > 10.2$ mJy). The percentage of AGN hosts in GOODS-S obtained in this case decreases 
to 40$\pm$4 per cent and is now consistent with the COSMOS one within 1$\sigma$ uncertainty. The overall fraction of $>99$ per cent significant AGN is 37 per cent, while it increases to 45 per cent and 55 per cent at $>95$ per cent and $>90$ per cent confidence levels, respectively. We label as AGN only those sources satisfying the \textit{F}-test at the $>99$ per cent confidence level, as this threshold is fairly conservative and reliable in terms of AGN contribution to the observed SED. In Fig. \ref{fig:fit_magphys}, the upper and middle panels represent sources with a $>99$ per cent significant AGN contribution, while that shown in the bottom panel has a less significant AGN contribution, hence classified as a ``purely star-forming galaxy''. 

We have verified that the star-forming galaxy SEDs adopted in the code were physically motivated. For instance, we found that our best-fit galaxy SEDs well reproduce the ``IR8'' relation, presented by \citet{Elbaz+11}, within $\pm$1$\sigma$ uncertainty. This finding also removes possible systematics related to the treatment of PAH emission component with respect to the total IR luminosity. We also checked the potential degeneracy between AGN and Starburst emission components when fitting the rest-frame mid-IR (3 $< \lambda/\mu$m $<$ 8) data points. Indeed, we note that the inter-stellar medium (ISM) emission in the {\sc magphys} star-forming SEDs includes dust and PAH emission down to 3$~\mu $m (see \citealt{DaCunha+08}). An excess above the pure stellar continuum will thus not be automatically ascribed to AGN. Besides, given the redshift distribution of our sources in Fig. \ref{fig:z_dist}, the rest mid-IR continuum is sampled at least by two filters in almost all observed SEDs (four IRAC bands, MIPS 24$~\
mu $m, and IRS 16$~\mu $m in GOODS-S). Such a rich spectral coverage allows us to better constrain the rest mid-IR emission through the available templates.

\subsubsection{AGN detection rate and evolution with $L_{1-1000}$}  \label{dependence}

The sources fulfilling the \textit{F}-test at $>$99 per cent confidence level are 1607/4343 (37 per cent). Our finding settles in an open debate concerning the fraction of AGN which are detected through independent selection methods. In the following we check the consistency between our results and the most recent estimates inferred from the analysis of different IR galaxy samples.

In the local Universe, \citet{Yuan+10} found the percentage of sources hosting an optically identified AGN to depend on the total infrared luminosity (8-1000$~\mu$m)\footnote{For historical reasons, throughout the paper the expression ``IR luminosity'' (or $L_{\rm IR}$) will refer to the luminosity integrated over the spectral range 8--1000$~\mu $m. If calculated in a different range (e.g. 1--1000$~\mu $m) it will be explicited.}. This fraction ranges from 20--30 per cent up to $\sim$60 per cent in local Luminous ($L_{\rm IR}\sim$10$^{11}$ L$_{\odot}$) and Ultra-Luminous ($L_{\rm IR}\sim$10$^{12}$ L$_{\odot}$) Infrared Galaxies (LIRGs and ULIRGs), respectively. Consistent results have been obtained by \citet{Imanishi+10} and \citet{Alonso-Herrero+12} through \textit{Akari}-IRC and \textit{Spitzer}-IRS spectroscopy of local LIRGs and ULIRGs. These authors independently found that the AGN detection rate increases with increasing $L_{\rm IR}$, from 25 per cent (LIRGs) up to 70 per cent (brightest ULIRGs), 
strenghtening the previous results of \citet{Lutz+98} based on \textit{Infrared Space Observatory} (ISO) data (from $\sim$15 per cent at $L_{\rm IR}<2\times$10$^{12}$ L$_{\odot}$ up to 50 per cent for brighter ULIRGs). The spectral decomposition has also been useful to effectively identify AGN: through detection of IR fine-structure lines, PAH equivalent widths and mid-IR continuum, several works (e.g. \citealt{Sajina+09}; \citealt{Nardini+10}; \citealt{Sajina+12}) claimed the presence of a significant AGN in 60--70 per cent of local and distant ULIRGs, as well as evident signatures of coexisting star formation.

At higher redshifts, the situation is more controversial. \citet{Olsen+13} analysed a mass-selected sample at $z\sim$2 within the \textit{Chandra Deep Field-South} (CDF-S) and ended up with a high fraction (43--65 per cent) of star-forming galaxies likely hosting an AGN, according to the X-ray classification taken from \citet{Xue+11}. \citet{Pozzi+12} studied a sample of 24 ULIRGs at $z\sim$2 (see \citealt{Fadda+10}) within the GOODS-South field and selected to have faint 24$~\mu$m flux densities ($0.14 < S_{24\mu m} {\rm / mJy}< 0.55$). They ended up with a smaller fraction ($\sim$35 per cent) of ULIRGs featuring signatures of AGN activity and claimed that their IR luminosity emission is dominated by starburst processes. 

\begin{figure}
\begin{center}
    \includegraphics[width=\linewidth]{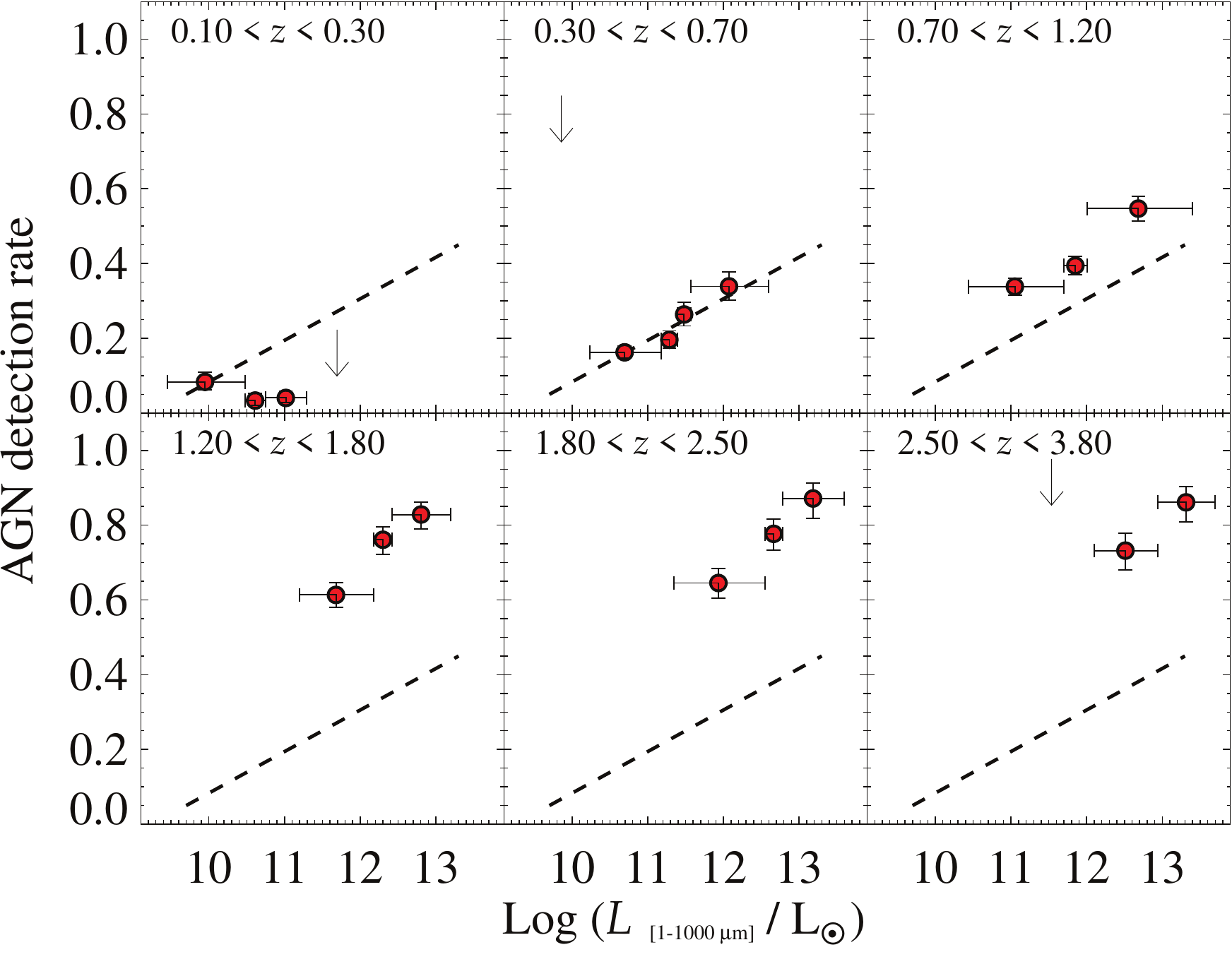}
\end{center}
 \caption{AGN detection rate in the 160$~\mu$m PEP-selected sample, as a function of the total IR (1--1000$~\mu$m) luminosity and redshift. In each redshift bin, the sample has been split into similarly populated luminosity bins. Error bars along the \textit{y}-axis correspond to 1$\sigma$ confidence limits, while downward arrows show 2$\sigma$ upper limits. The dashed line is a linear fit to the data of the second redshift bin to highlight the evolution of the AGN detection rate with redshift.}
   \label{fig:det_rate}
\end{figure}
  
In this work the sample has been split into six redshift bins (0.1$\leq$\textit{z}$<$0.3; 0.3$\leq$\textit{z}$<$0.7; 0.7$\leq$\textit{z}$<$1.2; 1.2$\leq$\textit{z}$<$1.8; 1.8$\leq$\textit{z}$<$2.5 and 2.5$\leq$\textit{z}$\leq$3.8) and into different IR luminosities. We check for the evolution of the AGN detection rate with total (AGN+starburst) 1--1000$~\mu$m luminosity $L_{\rm 1-1000}$ at different redshifts, as shown in Fig. \ref{fig:det_rate}. Error bars along the \textit{y}-axis show the $\pm$1$\sigma$ uncertainty (following  \citealt{Gehrels86} and \citealt{Gerke+07}), while the error bars along the \textit{x}-axis set the bin width. Downward arrows represent 2$\sigma$ upper limits to the expected AGN detection rate when only ``inactive'' galaxies are detected in that bin. We find an overall increase of the AGN detection rate as a function of $L_{\rm 1-1000}$, except in the lowest redshift bin. At $z<0.3$ the AGN hosts are mainly ``normal'' ($L_{\rm 1-1000} \leq 10^{11} \rm L_{\odot}$) star-forming 
galaxies. At this redshifts, the average AGN detection rate is $\sim$ 5--10 per cent and we do not find a clear trend with 1--1000$~\mu $m luminosity. This is likely due to the fact that the comoving volume covered by COSMOS is not large enough to detect sources with $L_{\rm 1-1000} \geq 10^{12} \rm L_{\odot}$.
 
At higher redshift (0.3$<$\textit{z}$<$3.8), the AGN detection rate does increase with increasing $L_{\rm 1-1000}$, which is consistent with results from previous studies. This finding might suggest that the probability to find AGN in FIR-selected galaxies increases as a function of $L_{\rm 1-1000}$. Besides the trend with $L_{\rm 1-1000}$, we point out the evidence for redshift evolution (see Fig. \ref{fig:det_rate}). Indeed, the AGN detection rate of $L_{\rm 1-1000} \sim 10^{11} \rm L_{\odot}$ does increase from 5 to 30 per cent up to $z\sim$1; galaxies with $10^{11} \leq L_{\rm 1-1000} / \rm L_{\odot} \leq 10^{12} $ host AGN activity with a probability rising from 25 per cent at $z\sim$0.5 to 60 per cent at $z\sim$2. Brighter ($L_{\rm 1-1000} \ge 10^{12} \rm L_{\odot}$) IR galaxies show an AGN detection rate rising from 55 per cent at $z \sim 1$ to 70--80 per cent at $z\sim$2--3. 

We also checked the potential dependence of this finding on selection effects. Indeed, our selection at 160$~\mu$m is certainly sensitive to warmer dust with increasing redshift, which might favour the detection of galaxies hosting AGN activity in more distant sources. However, given the redshift distribution of our \textit{Herschel}-selected sample in the GOODS-S and COSMOS field, we detected 35 objects only at $3 < z < 3.8$, where our selection wavelength spans the spectral range $35 < \lambda < 45 ~ \mu$m rest-frame. Moreover, the AGN templates adopted in this work (from \citealt{Feltre+12}; \citealt{Fritz+06}) typically peak around 15--20$~\mu$m (rest-frame), going down at longer wavelengths. We examined for each source the fractional energy contribution at 160$~\mu$m (observed-frame) due to AGN emission, as returned from the {\sc magphys+agn} SED-fitting analysis. We found the selection wavelength being poorly related to AGN activity, since its energy contribution to the total energy emitted 
at 160$~\mu$m is negligible ($<1$ per cent) with respect to that ascribed to star-formation, apart from a few exceptional cases (about 20 sources) that are generally characterised by an AGN dominated SED in the mid-IR. This check makes us reasonably confident that the selection at 160$~\mu$m does not affect the overall increase of the AGN detection rate with redshift.

\subsubsection{Comparison with mid-IR colour-colour AGN selection}  \label{agn_diagnostics}

\begin{figure}
\centering
     \includegraphics[width=\linewidth]{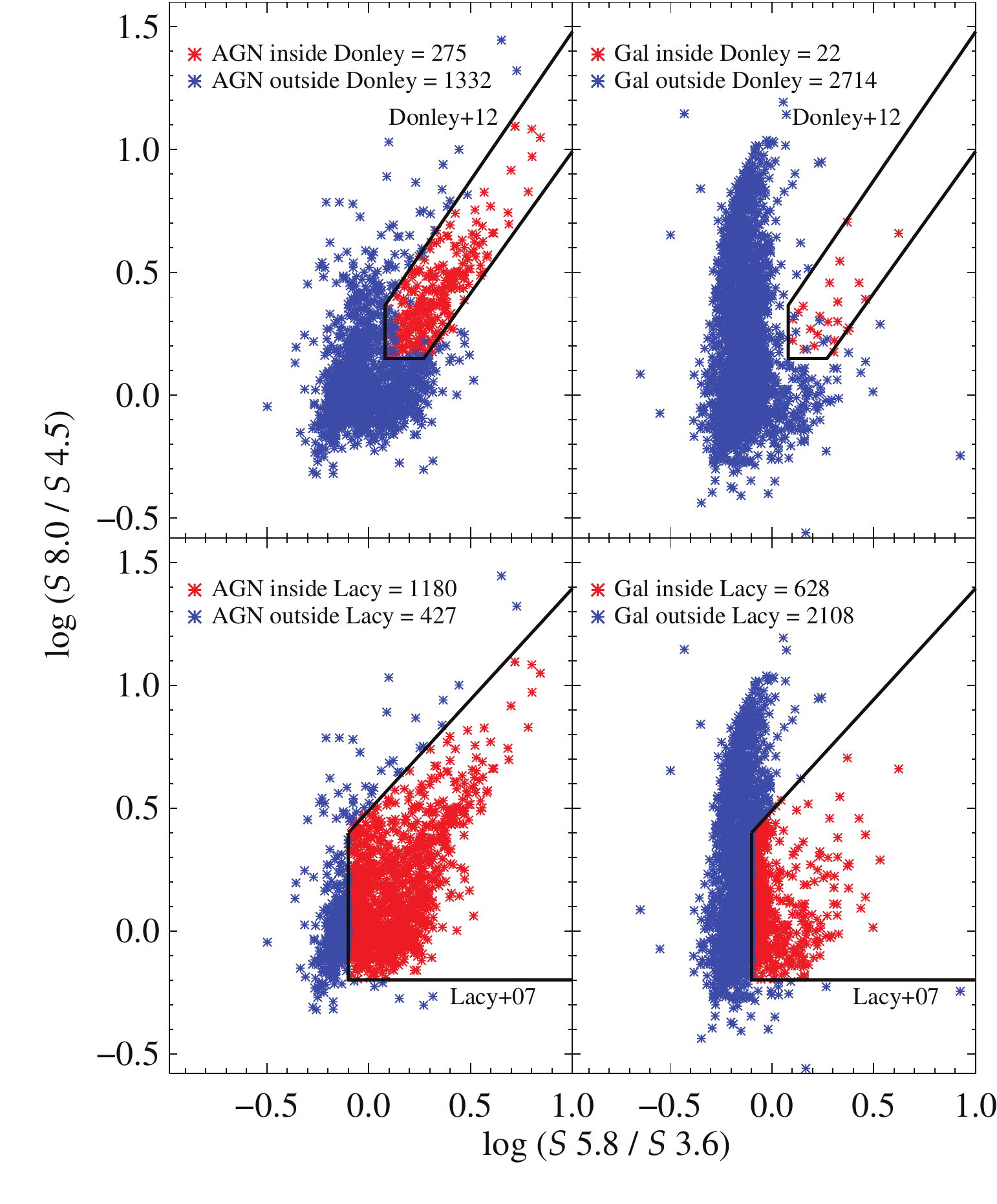}
     
 \caption{Comparison with AGN colour-colour selection criteria. \textit{Upper panels}: comparison with \citet{Donley+12}. \textit{Bottom panels}: comparison with \citet{Lacy+07}. Left and right panels refer to PEP sources classified as AGN and galaxies, respectively. Sources satisfying the corresponding criteria are marked with red asterisks, otherwise with blue asterisks.}
   \label{fig:agn_cc}
\end{figure}

In order to test the robustness of the method that we apply to estimate the AGN contribution in FIR-selected galaxies, we compare our findings with those obtained from an independent AGN diagnostic, specifically the mid-IR colour-colour selection. This criterion was first introduced by \citeauthor{Lacy+04} (\citeyear{Lacy+04}, \citeyear{Lacy+07}) by making use of \textit{Spitzer}-IRAC colours to isolate the AGN candidates, both obscured and unobscured. However, as already stated by \citet{Lacy+07}, a sample of so-called ``AGN candidates'' might suffer from a significant contamination from ``pure'' galaxy SEDs with similar mid-IR colours. More recently, \citet{Donley+12} revisited the previous method to provide a more reliable AGN selection criterion, although it was less complete at low X-ray luminosities ($L_{\rm [0.5-8]} < 10^{44}$ erg s$^{-1}$). In Fig. \ref{fig:agn_cc} we compare our classification with that derived from these mid-IR colour-colour diagnostics. We report the comparison with \citet{
Donley+12} in the upper panels and with \citet{Lacy+07} in the bottom ones. Left and right panels refer to our AGN and galaxy samples, respectively. Red asterisks mark the population of sources ``inside'' the corresponding wedge, whereas blue asterisks represent the sources in the PEP sample not satisfying the colour-colour criterion. 

As shown in Fig. {\ref{fig:agn_cc}}, some PEP sources, although inside the black solid wedge of \citet{Donley+12}, are labelled as ``outside''. This is why, in addition to the requirement of being inside the wedge, the IRAC colours of each source have to fulfill further conditions on IRAC fluxes at the same time to satisfy the Donley criteria. \footnote{Given the monochromatic fluxes ($S_{\lambda}$) at 3.6, 4.5, 5.8 and 8.0$~\mu $m, the criteria established by \citet{Donley+12} set the following additional conditions: $S_{3.6} < S_{4.5} < S_{5.8} < S_{8.0} ~$ at the same time, which make the black solid wedge a multi-dimensional region. }

The colour-colour wedge defined by the Donley criteria (black solid line) includes $\sim$17 per cent of our AGN sample and only $\sim$0.8 per cent of our galaxy sample. According to our classification, we confirm that \citet{Donley+12} selection is very efficient in avoiding contamination from purely star-forming galaxies. The overlapping fraction of AGN obviously increases when considering the wider \citet{Lacy+07} region: we find that 73 per cent of our PEP AGN and 23 per cent of our PEP galaxies are inside the wedge.

We checked the average integrated properties of PEP AGN inside or outside both colour-colour AGN selection criteria. As expected, objects satisfying the Donley criterion are mostly AGN-dominated systems in the mid-IR (both obscured and unobscured) and populate the highest luminosity tail of different luminosity distributions. Indeed, typical values for the 1--1000$~\mu $m AGN luminosity reach $\sim$10$^{12} \rm L_{\odot}$, as well as for the 1--1000$~\mu $m luminosity due to star formation. Moving towards the area delimited by the \citet{Lacy+07} criteria and outside the \citet{Donley+12} selection, the typical properties become less and less extreme, with Seyfert-like SEDs (i.e. dominated by the host galaxy light in the IR) being more common than QSO-like ones. 

In summary, we find the colour-colour criterion by \citet{Lacy+07} to be in reasonably good agreement with our classification and to represent an acceptable compromise between reliability and completeness. In addition, the Lacy wedge effectively rules out sources with 1--1000$~\mu $m SF luminosity $L_{\rm 1-1000} < 10^{11} \rm L_{\odot}$, where the AGN contribution to the IR becomes negligible ($\sim$ few  per cent) and also the probability to pick up AGN is generally lower (see \S~\ref{dependence} and Fig. \ref{fig:det_rate}).

PEP AGN as classified through SED decomposition occur in different regions with respect to PEP Galaxies in the mid-IR colour-colour diagram. This, together with the percentage of our PEP AGN and galaxies fitting with the Donley and Lacy criteria discussed above, further supports the robustness of our classification.

\begin{figure}
\begin{center}
    \includegraphics[width=\linewidth]{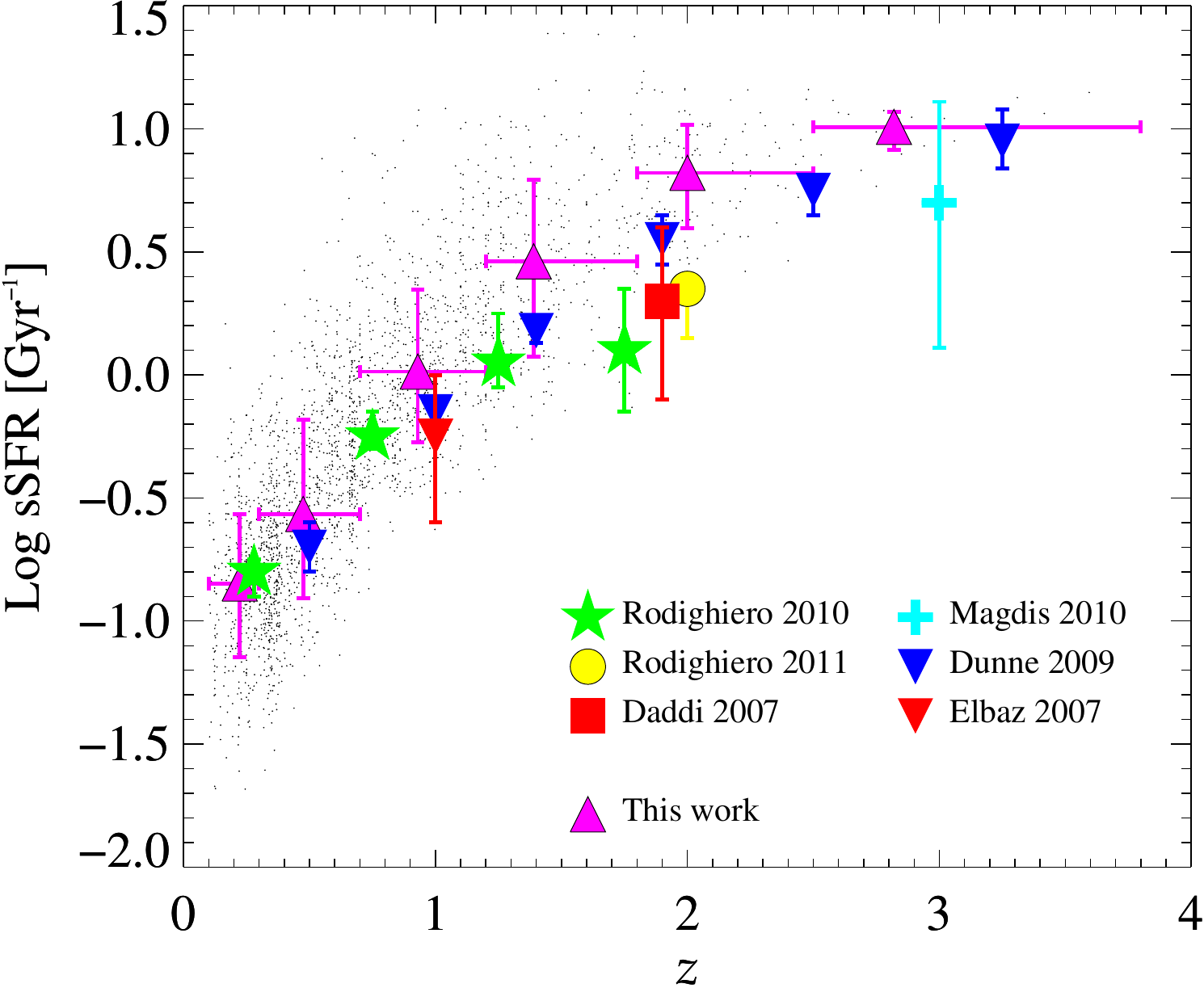}
\end{center}
 \caption{ Specific SFR as a function of redshift in the stellar mass range $10 < M_{\star} < 11$. Black dots represent the sSFR obtained through our SED decomposition for the joint (GOODS-S and COSMOS) sample. Magenta triangles set the median sSFR in each redshift bin. Horizontal error bars set the bin-width; vertical error bars enclose the $\pm$1$\sigma$ population of sources in each redshift bin. Various estimates from the literature, complemented with their respective error bars, are reported for comparison in the same stellar mass range.}
   \label{fig:ssfr}
\end{figure}
\subsubsection{Specific-SFR vs redshift}  \label{ssfr}

Our SED-fitting analysis allows us to get both stellar mass ($M_{\star}$) and SFR for each source. Therefore, to derive the specific star-formation rate (sSFR) for the entire \textit{Herschel}-selected sample and to infer its evolution with redshift.

Previous studies have highlighted the presence of a tight correlation between stellar mass and SFR in the local Universe (e.g. \citealt{Peng+10}, \citeyear{Peng+12}). These parameters trace the well-known ``main-sequence'' (MS) of star-forming galaxies. Such relation has been extended to $z\sim$1 (\citealt{Elbaz+07}) and up to $z\sim$2 (\citealt{Daddi+07}; \citealt{Rodighiero+11}) and $z\sim$3 (\citealt{Magdis+10}). In particular, \citet{Rodighiero+11} argued the presence of a bimodal regime in star-formation between galaxies lying on the MS and those on the off-sequence (i.e. $>$0.6 dex above the MS), with the former fueled through steady star-formation events, whereas the latter are experiencing enhanced and starbursting star-formation.

In this subsection, we compare the mean evolutionary trend of the sSFR with redshift with independent findings from the literature. The stellar mass is inferred from the best-fit template of the {\sc magphys+agn} SED-fitting. The SFR has been derived by converting the infrared (rest 8-1000$~\mu$m) luminosity through a standard \citet{Kennicutt98} relation, rescaled to a Chabrier IMF. In case of ``active'' galaxy, the AGN contribution has been subtracted from the total IR luminosity, otherwise the infrared emission is supposed to be due to star-formation only.

In Fig. \ref{fig:ssfr} we show the overall evolution of the sSFR with redshift for sources with $10 < M_{\star} < 11$. The joint (GOODS-S and COSMOS) sample has been split in six different redshift bins, as previously done in Fig. \ref{fig:det_rate}. Black dots represent each source of our sample, while magenta triangles show the median sSFR in each redshift bin. Error bars cover the $\pm$1$\sigma$ population in each \textit{z}-bin along the vertical axis and set the bin-width along the horizontal axis. In this plot only sources above the nominal SFR limit have been considered. The cut in SFR at a given redshift bin corresponds to the minimum SFR value spanned by the SFR-$z$ distribution at the upper bound of the same $z$-bin. This cut ensures to get a complete sample in SFR in each $z$-bin.

For comparison, in Fig. \ref{fig:ssfr} we report other independent estimates of the sSFR as a function of redshift. \citet{Rodighiero+10} studied an \textit{Herschel}-selected sample in the GOODS-North field, while \citet{Rodighiero+11} extended the previous analysis with \textit{Herschel} data to the COSMOS and the GOODS-South field. \citet{Daddi+07} and \citet{Dunne+09} analysed a sample of \textit{k}-selected galaxies, in the GOODS-South and in the Ultra Deep Survey (UDS), respectively. \citet{Elbaz+07} carried on a IRAC-based work (3.6 and 4.5$~\mu$m selection) in the GOODS-South. \citet{Magdis+10} investigated a sample of IRAC-detected Lyman Break galaxies in the \textit{Hubble}-Deep Field North (HDF-N).

We find our trend to be in fair agreement with the other independent estimates, within 1$\sigma$ uncertainty. However, a more detailed analysis of the AGN contribution in the $M_{\star}$-SFR plane, for both detected and undetected X-ray sources, is postponed to a forthcoming paper.

\section{AGN Bolometric Luminosity Function}  \label{acc_lf}

In this section we investigate the evolution of the AGN bolometric luminosity function (LF) over the whole joint GOODS-S and COSMOS sample. Globally we collect 4343 objects, but just the fraction showing a substantial AGN contribution has been considered for the following analysis (37 per cent at 99 per cent CL). The sample of interest has been split into six different redshift bins (as done in Fig. \ref{fig:det_rate}) and 10 luminosity bins, from 10$^8$ up to 10$^{15} ~ \rm L_{\odot}$, with a logarithmic bin-width equal to 0.7.

\begin{figure*}
\begin{center}
    \includegraphics[width=180mm,keepaspectratio]{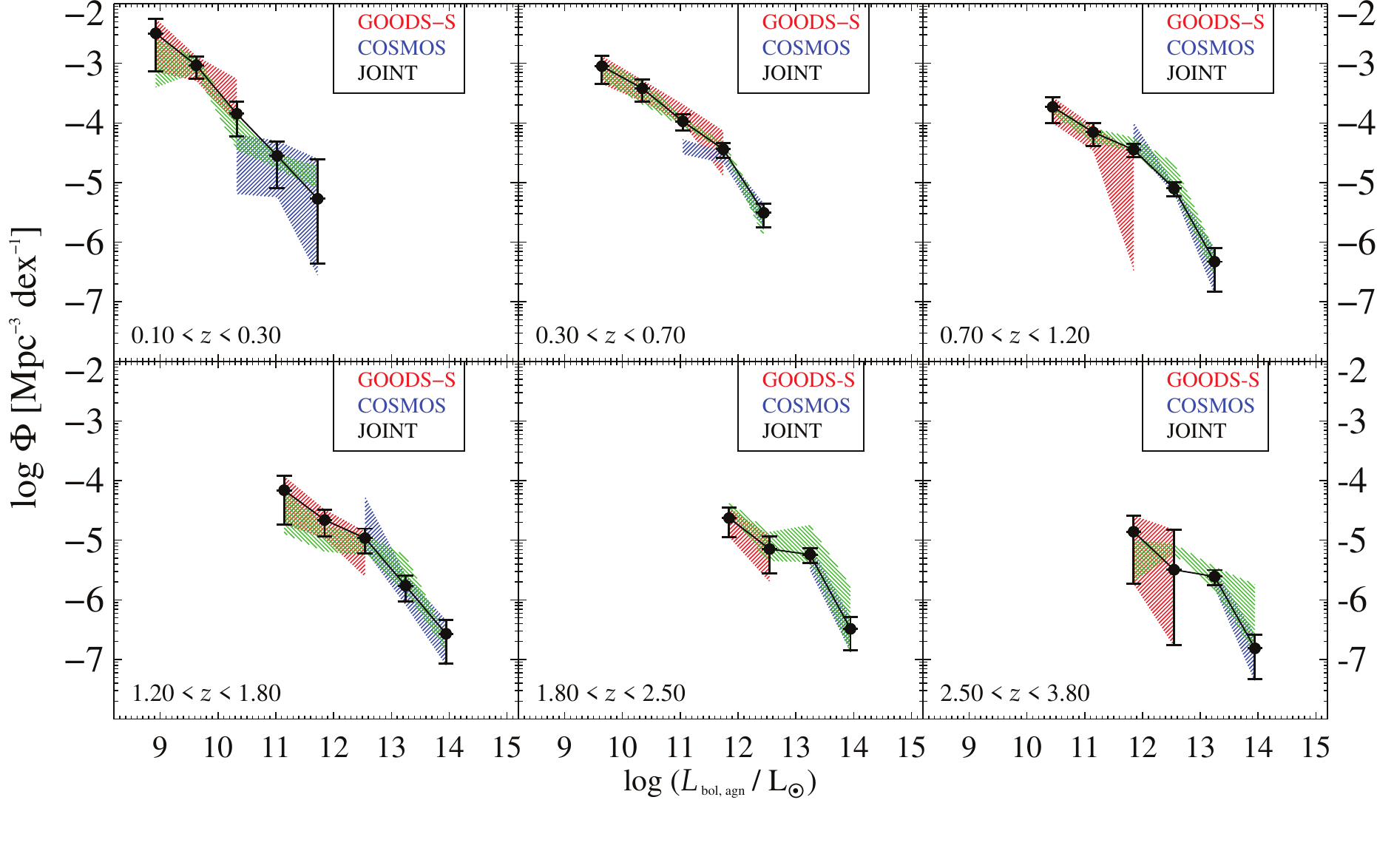}
\end{center}
 \caption{Accretion luminosity function (LF) in different redshift bins for the PEP sample. Red and blue shaded areas trace the $\pm$1$\sigma$ Poissonian uncertainties separately for GOODS-S and COSMOS observed data points, respectively. The joint (GOODS-S and COSMOS) LF is represented by black circles, with $\pm$ 1$\sigma$ uncertainties (black vertical bars) related to the joint sample. The green dashed areas show the $\pm 1 \sigma$ range of $\Phi_{\rm MC}(L,z)$ obtained through Monte Carlo simulations (see text for details).} 
   \label{fig:lf_data_fields}
\end{figure*}
%
\subsection{Method} \label{method_lf}

The computation of the LF relies on the non-parametric $1/V_{\rm max}$ method introduced by \citet{Schmidt68}, where $V_{\rm max}$ represents the maximum comoving volume that allows a source of a given luminosity to be observable above the flux limit of the survey. Since we consider sources coming from two different samples (GOODS-S and COSMOS), we use the method of \citet{Avni+80}, who extended the previous $1/V_{\rm max}$ method to coherently join and analyse simultaneously different samples. For each source of our sample we compute the $V_{\rm max}$ as follows:
\begin{equation}
V_{\rm max} = \int_{z_{\rm min}}^{z_{\rm max}}{ \frac{dV}{dz}} ~ \Omega(z) ~ dz
   \label{eq: vmax}
\end{equation}
with $z_{\rm min}$ being the lower boundary of a given redshift bin and $z_{\rm max}$ the minimum between the upper boundary of the redshift bin and the maximum redshift value available for the source of interest to be detected. Here $\Omega(z)$ represents the effective area of visibility and is computed as the fraction of the geometrical projected sky area $\Omega_{\rm geom}$, where the sensitivity of the instrument is enough to let the source be observable, according to the formula:
\begin{equation}
\Omega(z) = \Omega_{\rm geom} \cdot f_{\rm c} (z) ~ ,
   \label{eq: omega_z}
\end{equation}
where $f_{\rm c} (z)$ is the flux completeness correction obtained from simulations by \citeauthor{Berta+10} (\citeyear{Berta+10}, \citeyear{Berta+11}). 
For a given luminosity and redshift bin, we computed the luminosity function as follows:

\begin{equation}
\Phi(L,z) = \frac{1}{\Delta \log L} ~ \sum\limits_{i=1}^n ~ \frac{1}{V_{\rm max, i}}
\end{equation}
where $\Delta \log L$ is equal to 0.7, as mentioned before. The COSMOS field has a 93 per cent redshift completeness, regardless of PACS flux densities, so that we multiply $\Phi(L,z)$ in COSMOS by a factor of 1.07, whereas the GOODS-S field is 100 per cent complete in redshift and therefore no correction is needed. Moreover, we account for the fraction of the COSMOS sample without a MIPS-24$~\mu $m counterpart inside the MIPS area. Indeed, in the context of a multi-component SED-fitting, the lack of the 24$~\mu $m detection prevents the possible torus component from being constrained, resulting in low reliability best-fits. This corresponds to $\sim 3.5$ per cent of the PEP sample, roughly independent of redshift and AGN luminosity. We take into account their contribution by uniformly increasing $\Phi(L,z)$ by a factor of 3.5 per cent.

\begin{figure*}
\begin{center}
    \includegraphics[width=180mm,keepaspectratio]{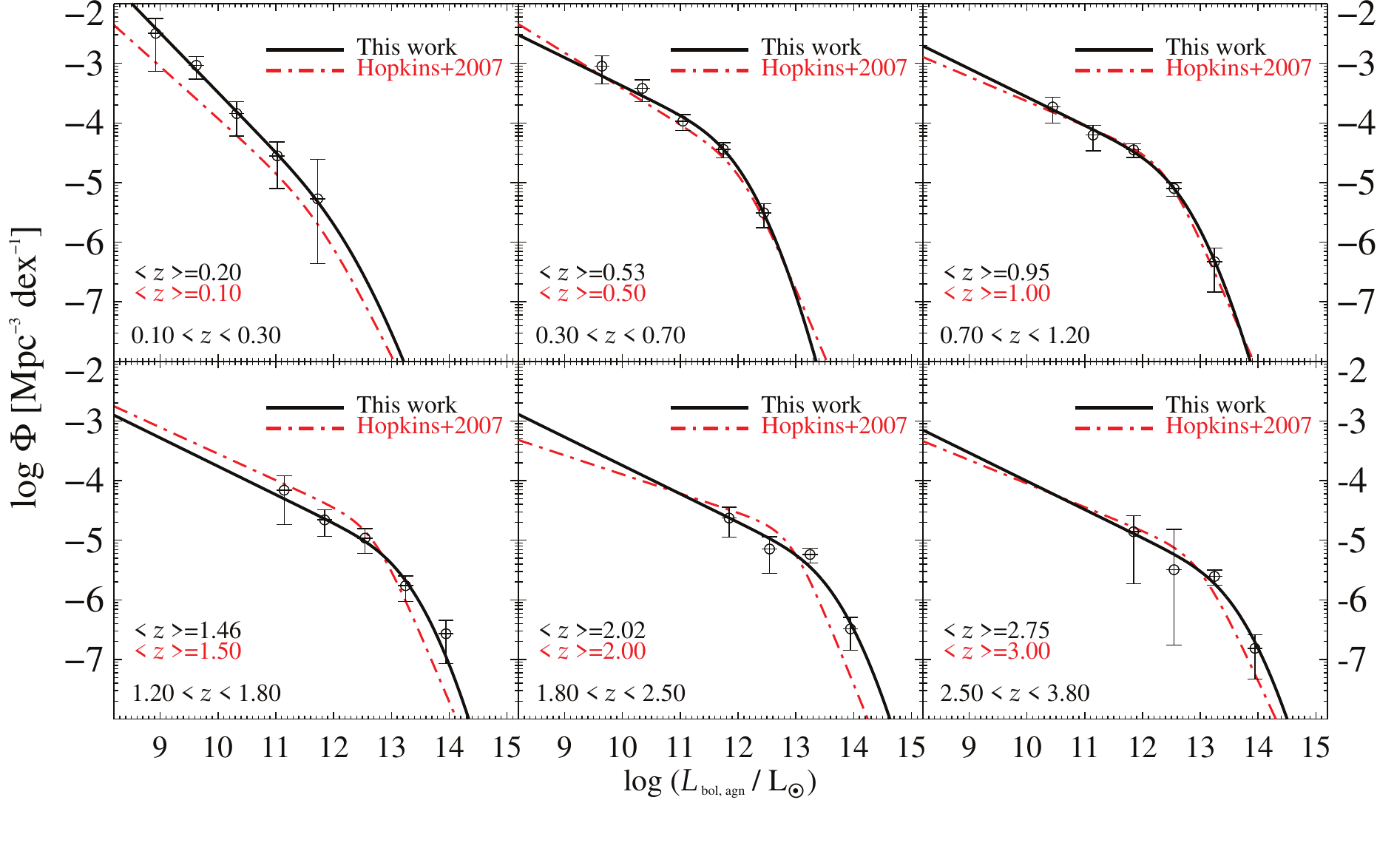}
\end{center}
 \caption{Accretion LF (open circles with $1\sigma$ Poissonian uncertainties) derived from PEP data. The black solid line represents our best-fit LF, whereas the red dash-dotted line is the best-fit curve of the AGN bolometric LF, as derived by \citeauthor{Hopkins+07} (\citeyear{Hopkins+07}, hereafter H07). In each redshift bin we report the median redshift value of the AGN population, both for this work and for H07.}
   \label{fig:lf_fit}
\end{figure*}

\subsection{AGN bolometric LF and its uncertainties}  \label{monte_carlo}

In Fig. \ref{fig:lf_data_fields} we show our accretion LF in different redshift bins. Error bars correspond to $\pm$ 1$\sigma$ Poissonian uncertainties (\citealt{Marshall85}), while in case of one single object in a given luminosity bin, we provide a 90 per cent uncertainty following the recipe of \citet{Gehrels86}. We computed separate LFs for GOODS-S and COSMOS (red and blue areas, respectively), and the total LF for the joint (GOODS-S {\sc +} COSMOS) sample (black circles). As expected, in all redshift bins the GOODS-S field extends down to lower luminosities, whereas COSMOS tipically samples higher luminosities.

\begin{table*}
   \caption{ List of best-fit parameters of the AGN bolometric LF with related $\pm 1 \sigma$ uncertainties. }
\begin{tabular}{lcccc}
\hline
$ ~~~~~ z$-bin & $\alpha$ & $\sigma$ & $\log_{10} (L^{\star} / \rm L_{\odot})$ & $\log_{10} \Phi^{\star}$ ( \rm Mpc$^{-3}$ \rm dex$^{-1}$) \\
\hline
$0.1 \leq z < 0.3$  & 1.97${\pm 0.25}$ & 0.54${\pm 0.31}$ & 10.91${\pm 1.96}$ & --4.35$+0.93$   \\
$0.3 \leq z < 0.7$  & 1.48${\pm 0.22}$ & 0.54${\pm 0.31}$ & 11.34${\pm 0.14}$ & --4.03${\pm 0.31}$  \\
$0.7 \leq z < 1.2$  & 1.48${\pm 0.22}$ & 0.54${\pm 0.31}$ & 11.96${\pm 0.15}$ & --4.47${\pm 0.31}$  \\
$1.2 \leq z < 1.8$  & 1.48${\pm 0.22}$ & 0.54${\pm 0.31}$ & 12.61${\pm 0.24}$ & --5.00${\pm 0.55}$  \\
$1.8 \leq z < 2.5$  & 1.48${\pm 0.22}$ & 0.54${\pm 0.31}$ & 12.96${\pm 0.18}$ & --5.15${\pm 0.44}$  \\
$2.5 \leq z \leq 3.8$  & 1.48${\pm 0.22}$ & 0.54${\pm 0.31}$ & 12.92${\pm 0.22}$ & --5.40${\pm 0.61}$  \\
\hline
\end{tabular}

\label{tab:fit_param}
\end{table*}
The AGN bolometric LF shown here has been already corrected for incompleteness in accretion luminosity. We briefly describe our method as follows and refer the reader to the Appendix \ref{appendix} for a more detailed explanation. Our approach to account for the incompleteness in accretion luminosity follows that presented by \citet{Fontana+04}, except for the parameters involved in such analysis: accretion power rather than stellar mass, and 160$~\mu $m flux $S_{\rm 160}$ rather than K-band flux. Since a not negligible fraction of active galaxies might be missed by our FIR-based selection, it is necessary to quantify such missed AGN population. Starting from a FIR-selected sample, the incompleteness in accretion luminosity might be evaluated by looking at the distribution of flux ratio between accretion flux $S_{\rm accr}$ (defined as the bolometric flux corresponding to a given $L_{\rm bol, AGN}$ and redshift) and $S_{\rm 160}$, traced by our AGN population. By assuming that this distribution does not 
change either with redshift or with \textit{Herschel} flux density, it is possible to shift that down in $S_{\rm accr}$ and to virtually sample relatively high values of $L_{\rm bol, AGN}$ in weakly star-forming galaxies, not observable by \textit{Herschel}. We iterate this shift down in $S_{\rm accr}$ as long as the expected number of missed AGN is equal to the number of observed AGN. The latter step identifies a threshold in accretion flux $S_{\rm accr, lim}$, corresponding to correction for incompleteness by a factor of two. The curve which parametrises the incompleteness is used to compute the maximum comoving volume where each source with $S_{\rm accr} \geq S_{\rm accr, lim}$, either detected or undetected by \textit{Herschel}, is expected to be placed. We carried out this analysis separately for GOODS-S and COSMOS fields. As mentioned before, the AGN bolometric LF shown in Fig. \ref{fig:lf_data_fields} already incorporates the correction for incompleteness. This means that our LF is supposed to account 
for all active galaxies with $S_{\rm accr} \geq S_{\rm accr, lim}$, either above or below the \textit{Herschel} detection limit. This allows us to remove the observational bias due to our FIR selection.

In the luminosity bins which are populated by both GOODS-S and COSMOS AGN at the same time, the two accretion luminosity functions are in reasonably good agreement one with the other. In the first redshift bin we find a large gap ($\sim$1 dex) in the connection point between the two surveys, but the very low statistics (only two objects per field) makes the two LF broadly consistent within 2$\sigma$.

A Monte Carlo (MC) simulation has been performed to account for all the uncertainties in $L_{\rm bol, AGN}$ associated to the objects in the same redshift and luminosity bin. For each object belonging to a given $(L_{\rm bol, AGN},z)$ bin, we randomly extract an individual $L_{\rm bol, AGN}$ value from the whole PDF distribution, using the intrinsic shape of the PDF as the weight. Then, using these random values of $L_{\rm bol, AGN}$, we compute the LF for this simulated sample. By iterating the same MC simulation 100 times, we characterise a range of $\Phi_{\rm MC}(L,z)$ defining the uncertainty region associated with each LF datapoint $\Phi(L,z)$. In Fig. \ref{fig:lf_data_fields} we show our accretion LF (black circles) with its related $\pm 1 \sigma$ confidence region (green dashed areas). The size of the uncertainty region is tipically comparable to the Poissonian error bars, except for the less populated bins (large Poissonian error bars). The limits drawn by MC simulations have been used to 
provide an uncertainty range for the integrated LF (see \S~\ref{bhad}).

\subsection{Evolution of the AGN bolometric LF}  \label{best_fit_lf}

We fit our set of data points making use of a modified version of the Schechter function (\citealt{Saunders+90}), as follows:
\begin{equation}
\Phi(L) \rm d\log L = \Phi^{\star}\left(\frac{L}{L^{\star}}\right)^{1-\alpha} \exp \left[-\frac{1}{2\sigma^2}\log_{10}^2\left(1+\frac{L}{L^{\star}}\right)\right] \rm d\log L
\end{equation}
which behaves as a power law for $L << L^{\star}$ and as a Gaussian in $\log L$ for $L << L^{\star}$. The couple of parameters ($\Phi^{\star}$, $L^{\star}$) represent the normalization and luminosity of the knee of the distribution, respectively. Two other parameters ($\alpha$, $\sigma$) are set to shape the low and high luminosity tails of the best-fit function.

\begin{figure}
\begin{center}
    \includegraphics[width=\linewidth]{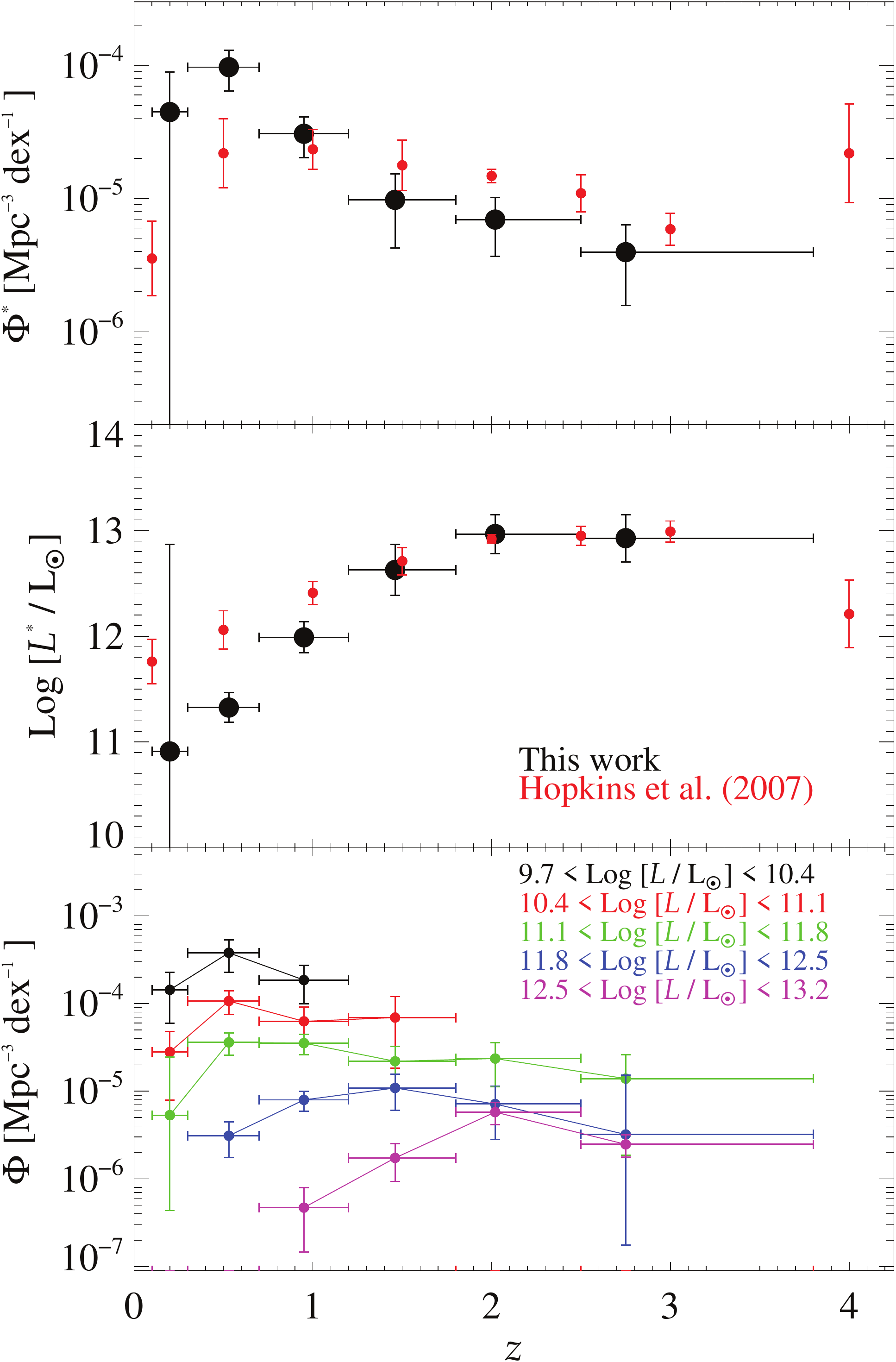}
\end{center}
 \caption{\textit{Top panel}: evolution with redshift of the normalization value $\Phi^*$, both for this work (black circles) and for H07 (red circles), taken from the respective best-fit LF. \textit{Middle panel}: evolution of the knee luminosity of the LF. Error bars correspond to the best-fit uncertainty in each redshift bin. \textit{Bottom panel}: number density of different luminosity bins as a function of redshift.}
   \label{fig:lf_par}
\end{figure}

We make use of a non-linear least square fitting routine to find the set of parameters that better reproduce the observed $\Phi(L,z)$ through a modified Schechter function. In Fig. \ref{fig:lf_fit} the best-fit functions have been overplotted on the data. The black line represents our best-fit, while the red dot-dashed line is the Quasar Bolometric LF as derived by \citeauthor{Hopkins+07} (\citeyear{Hopkins+07}, hereafter H07) from X-ray data. They used a wide compilation of observed AGN luminosity functions, from the mid-IR through optical, soft and hard X-ray data to build up a self-consistent AGN bolometric luminosity function. H07 adopted a distribution of luminosity dependent bolometric corrections for each selection wave band, as well as a large baseline of intrinsic (i.e. corrected for obscuration) AGN SED shapes (from \citealt{Richards+06}; \citealt{Steffen+06}). However, the intrinsic SEDs of \citet{Richards+06} also include an additional IR ($\lambda >$ 1$~\mu $m) contribution arising from the 
torus reprocessed radiation. This IR excess (with respect to a blackbody emission in the Rayleigh-Jeans regime) has to be removed to avoid double counting the input energy of the accretion disc. As already done by \citet{Merloni+13}, we independently evaluated such an excess to be of the order of $33$ per cent. As a consequence, all the AGN bolometric luminosities taken from H07 have been reduced by a factor $\sim$0.33 ($\sim$0.18 dex). We split our set of data points in order to sample similar mean redshifts to H07. 

There are four functional parameters controlling the Schechter function, but not more than five luminosity bins are sampled per redshift bin (i.e. there is either 0 or 1 degree of freedom). As a consequence, not all the best-fit parameters could be sufficiently constrained at the same time. For this reason, we choose to let all the functional parameters be free in the third redshift bin only, and to fix $\alpha$ and $\sigma$ at those resulting values. In the first redshift bin, the $\alpha$ value was allowed to vary, but $\sigma$ was fixed. 

\begin{figure*}
\centering
     \includegraphics[width=120mm,keepaspectratio]{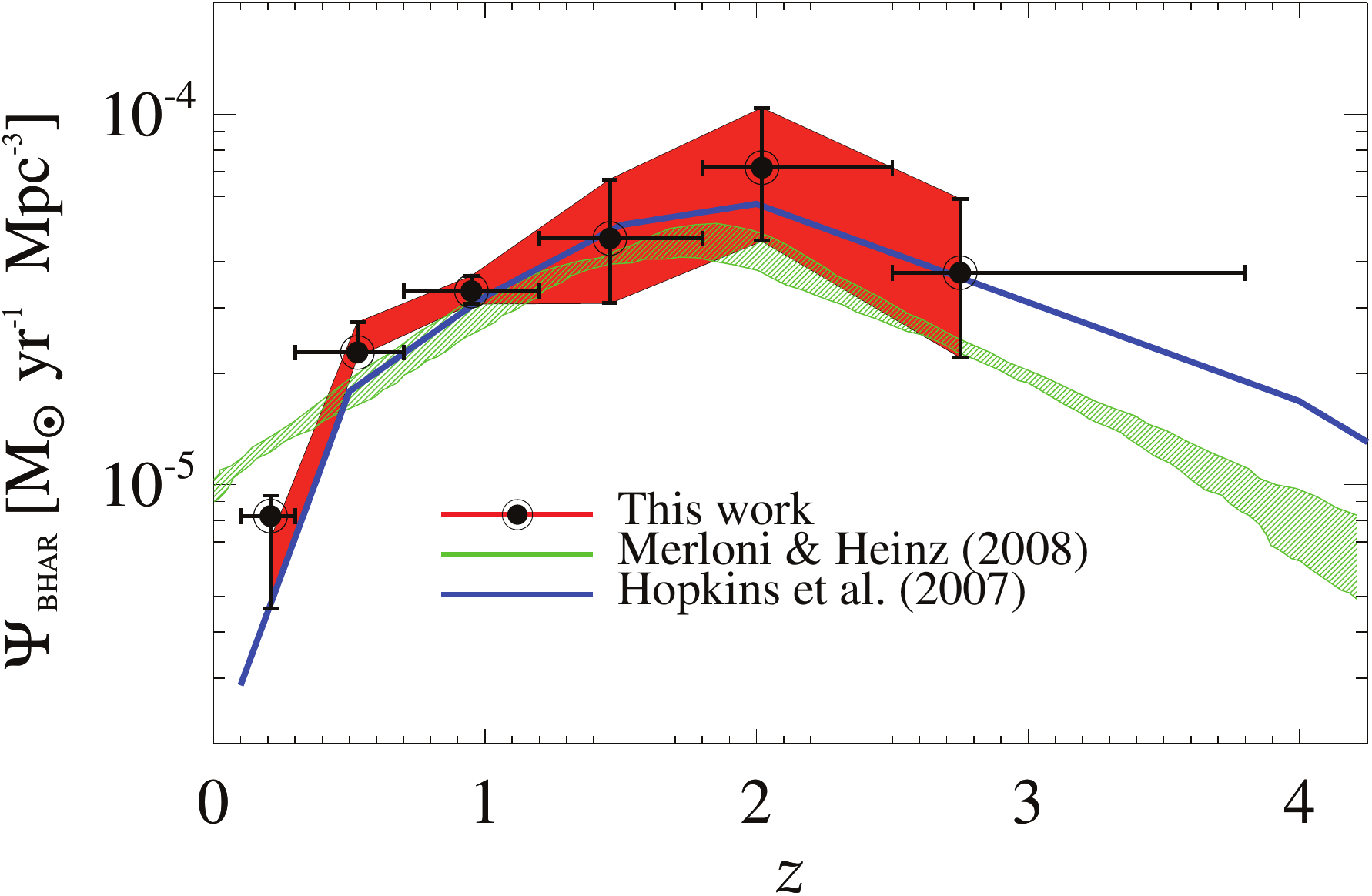}

\caption{Black Hole Accretion Rate Density estimate from the AGN bolometric LF as a function of redshift (black circles). The red shaded area shows the $\pm~1\sigma$ uncertainty region. Previous estimates from different selection wavelengths (from \citealt{Merloni+08}, and H07) are reported for comparison. }
   \label{fig:bhad}
\end{figure*}

This configuration implies that the shape of the Schechter function is defined in the first and third redshift bins and is allowed to scale rigidly along the (x,y) axes to fit the observed LF in the other redshift bins.

In Table \ref{tab:fit_param} we list all the best-fit parameters achieved through the least squares fitting, together with their related $\pm 1\sigma$ errors. Since only one object populates the highest luminosity bin of the first redshift slice, the resulting best-fit does not allow us to constrain either $\Phi^{\star}$ or $L^{\star}$ (see Table \ref{tab:fit_param}). 

Despite the uncertainties characterising the first \textit{z}-bin, there is evidence in favour of an evolution of the parameters ($\Phi^{\star}$, $L^{\star}$) as a function of redshift. As show in Fig. \ref{fig:lf_par} (top panel), the normalization value has a maximum around $z\sim$0.5, then it decreases to $z\sim$3; also H07 do find a similar evolution of $\Phi^*$ with redshift. In the middle panel, the knee of the modified Schechter LF increases with redshift, at least up to $z\sim$2.  Comparing the LF of our work with Table 2 of H07 \footnote{Note that the $L^{\star}$ values derived by H07 have been corrected in this work by a factor of $\sim$0.18 dex.}, the knee of the LF is systematically at lower luminosities that what was inferred by H07 in the respective redshift bins, except at z$>$1.5, where our $L^{\star}$ estimates are consistent with H07. This is likely due to the fact that we are biased against early-type (and/or weakly star-forming) galaxies, which are known to host the most luminous AGN in 
the local Universe. Letting the number density of early-type galaxies decrease towards higher redshift, one might also expect that fewer and fewer passive galaxies host an active SMBH, as predicted by the major merger evolutionary scenario (e.g. \citealt{Sanders+88}). This would imply that the bulk of the SMBH growth would be progressively dominated by late-type galaxies, so that at higher redshift our $L^{\star}$ values become consistent with those found by H07. In Fig. \ref{fig:lf_par} (bottom panel) we show the evolution with redshift of the AGN number density, for different luminosity bins. As already pointed out from several works (e.g. \citealt{Hasinger+05}; \citealt{Silverman+08}) from X-ray data, we confirm an anti-hierarchical growth for SMBHs harboured in star-forming galaxies selected by \textit{Herschel}, with the most luminous ones evolving faster and peaking at higher redshift.

\section{SMBH growth across cosmic time}  \label{bhad}

Given the estimated LF, we are able to derive the Black Hole Accretion Rate Density (BHAD or $\Psi_{\rm bhar}$) over cosmic time, from $z\sim$3 down to the present epoch. This quantity is fundamental for characterising the effective growth of AGN and is defined by the following expression:
\begin{equation}
\Psi_{\rm bhar}(z) = \int_{0}^{\infty}{ \frac{1-\epsilon_{\rm rad}}{\epsilon_{\rm rad} ~ c^2} ~ L_{\rm bol, AGN} ~ \phi(L_{\rm bol, AGN}) ~ d\log L_{\rm bol, AGN}}
\end{equation}
where all the ingredients are already available. The only parameter that still needs to be adopted is the radiative efficiency. We use $\epsilon_{\rm rad}$  = 0.1 and constant with redshift and intrinsic AGN luminosity (see H07).

\subsection{Comparison with previous results}  \label{comparison}

In Fig. \ref{fig:bhad} we report our BHAD estimate compared with previous findings from the literature.

\citeauthor{Merloni+08} (\citeyear{Merloni+08}, hereafter MH08) estimated the BHAD from the evolution of the hard X-ray LF taken by \citet{Silverman+08} and accounting for a distribution of luminosity-dependent $N_{\rm H}$ (Hydrogen column density along the line of sight) and a set of X-ray bolometric corrections from \citet{Marconi+04}. By solving the continuity equation for the SMBH mass function (assuming its local value from \citealt{Shankar+09}) and combining that with the evolution of the hard X-ray LF, they were able to trace the BHAD also as a function of black hole mass and accretion rate. 
It is worth mentioning that from Fig. 4 of MH08 it is shown that most of the local ($z\sim$0.1) SMBH growth is hidden in radiatively inefficient accreting systems, whose released energy is dominated by kinetic feedback rather than radiative losses. Such ``silent'' (Eddington ratio $\lambda_{\rm Edd} < 3 \times 10^{-2}$) AGN population probably does not enter our sample and could be responsible for the discrepancy (by a factor of about 1.7) shown in Fig. \ref{fig:bhad} in the local BHAD between MH08, H07 and our estimate. As mentioned in \S~\ref{bol_info}, H07 built up an observed quasar LF by consistently connecting different LFs, each one computed in a single wave band.

To guarantee a coherent comparison between our study and other works taken from the literature, we checked that both MH08 and H07 have accounted for the fraction of missed AGN because of their own selection effects. MH08 estimated the fraction of highly-obscured ($N_{\rm H} > $ 1.5$\times 10^{24}$ cm$^{-2}$) AGN from the X-ray background (XRB) sythesis models by \citet{Gilli+07}. In particular, they calculated the expected fraction between observed and unabsorbed 2--10 keV X-ray luminosity, by assuming a canonical AGN X-ray spectrum (absorbed power-law with photon index $\Gamma$=1.8; \citealt{Tozzi+06}), and as a function of obscuration in the range $ 21 < \log (N_{\rm H} / \rm cm^{-2}) < 24$. For $\log (N_{\rm H} / \rm cm^{-2}) >$ 24 they assumed that the ratio between observed and intrinsic X-ray luminosity was of the order of 2 per cent (\citealt{Gilli+07}).

H07 estimated the incompleteness fraction with a set of luminosity dependent $N_{\rm H}$ distributions. First, they collected  a wide set of luminosity-dependent AGN spectral shapes and calculated the bolometric correction distributions, in case of no obscuration. In addition, they implemented three possible models for obscuration. For each parametrization, H07 calculated the expected amount of extinction in X-rays, optical and mid-IR. Finally, through a convolution between AGN templates, bolometric correction distributions and $N_{\rm H}$ distributions, they were able to predict the incompleteness fraction of AGN in each band.

Our estimates (black circles) have been derived by integrating the best-fit curve of the LF, down to 10$^8 \rm L_{\odot}$, in each redshift bin. The red shaded area represents the $\pm$ 1$\sigma$ uncertainty region. The latter has been computed by accounting for the 100 different integrated $\Phi_{\rm MC}(L,z)$ values and plotting their cumulative distribution at 16 and 84 percentiles in each redshift bin. The BHAD seems to evolve quickly from $z>3$ to $z\sim 2$, where it shows a peak, then decreases towards the present epoch. Despite our AGN sample having been selected and analysed independently of the previous ones, the overall trend is consistent with others taken from the literature, as shown in Fig. \ref{bhad}. This is not surprising, as both our estimate and that obtained by H07 and MH08 have been already corrected for their own incompleteness effects and should be in fair agreement as a proof of their mutual consistency. All the estimates are plotted for the canonical case $\epsilon_{\rm rad}=$0.1. A 
different value of $\epsilon_{\rm rad}$ simply results in a change of the BHAD normalization.

\begin{figure}
\centering
     \includegraphics[width=\linewidth]{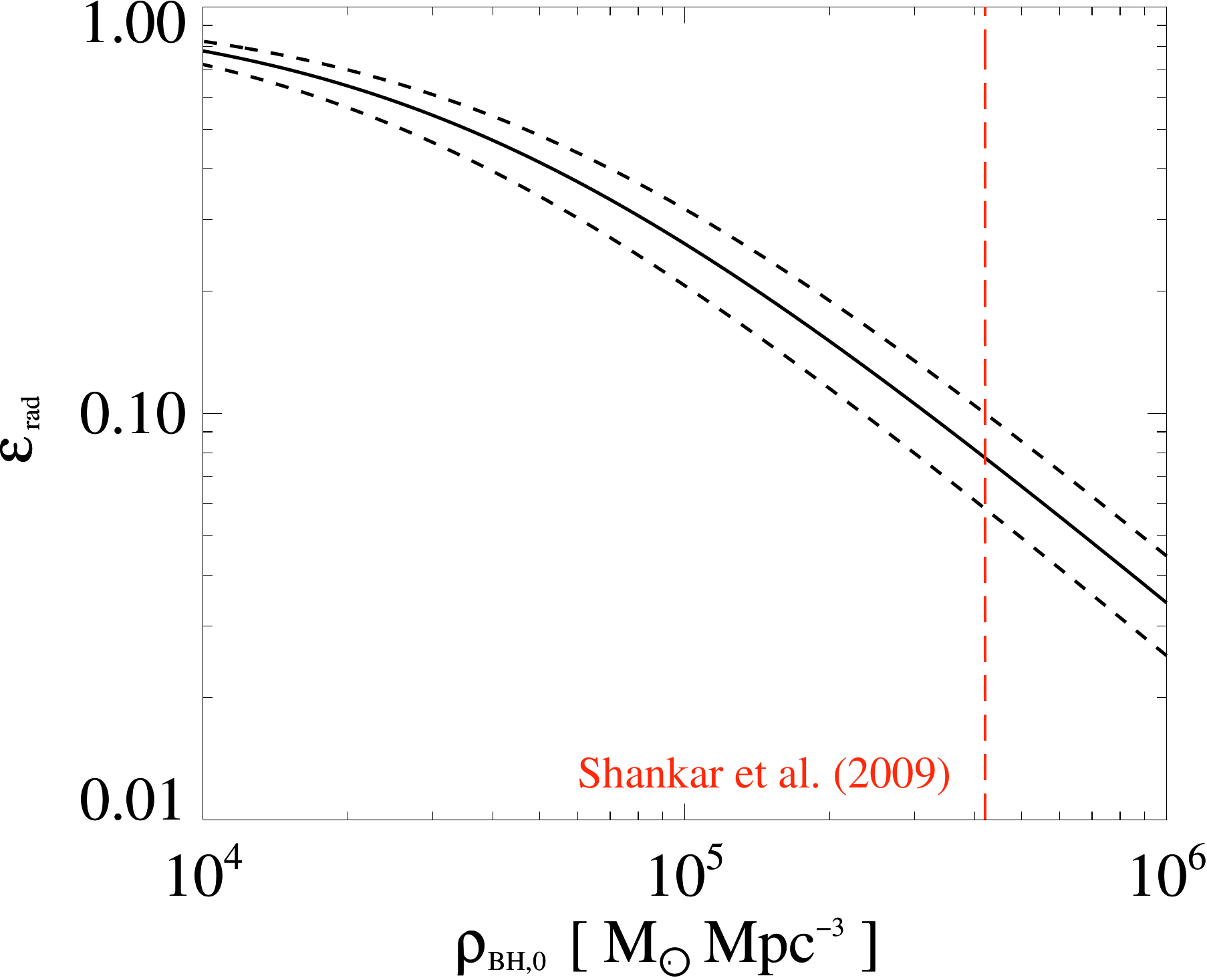}
     
\caption{SMBH radiative efficiency $\epsilon_{\rm rad}$ as a function of the local BH mass density predicted by this work (black solid line, with $\pm$ 1$\sigma$ uncertainty shown by the dashed lines). The vertical red dashed line is the estimate of the observed local BH mass density by \citet{Shankar+09}. }
   \label{fig:bh_mass_density}
\end{figure}

An important constraint that allows us to test the consistency of the integrated BHAD$(z)$ is to deal with Soltan's argument, i.e. deriving the local BH mass density and comparing it to the estimate achieved from observations. This represents the overall energy density released by SMBHs during cosmic time, down to $z=0$. One of the most recent estimates comes from \citet{Shankar+09}, as mentioned before. Tuning the integrated BHAD evolution in order to reproduce the observed $\rho_{\rm bh, 0}$ is a way to check the range spanned by the radiative efficiency $\epsilon_{\rm rad}$. From the standard relativistic accretion theory it is known that $\epsilon_{\rm rad}$ = 0.06--0.20 (\citealt{Novikov+73}). We integrate the BHAD over the redshift range from \textit{z}=6 to \textit{z}=10$^{-5}$ (but taking \textit{z}=$\infty$ as upper boundary would change our result by a factor $<1 $ per cent). By assuming $\epsilon_{\rm rad}=$0.1, as shown in Fig. \ref{fig:bh_mass_density}, our local BH mass density is
\begin{equation}
\rho_{\rm bh, 0} = 3.1^{+1.0}_{-0.8} ~\times ~ 10^5 ~ \rm M_{\odot} \rm Mpc^{-3}
\end{equation}
which is consistent with previous estimates. Indeed, for a typical matter-to-radiation conversion efficiency $\epsilon_{\rm rad}$ = 0.1, \citet{Yu+02} found $\rho_{\rm bh, 0}$ = 2.9$\pm 0.5 \times$10$^5~\rm M_{\odot} ~ \rm Mpc^{-3}$; \citet{Marconi+04} reported $\rho_{\rm bh, 0}$ =(4.6$^{+1.9}_{-1.4}$) $\times$ 10$^5\rm M_{\odot} ~ \rm Mpc^{-3}$; \citet{Shankar+09} achieved a value of $\rho_{\rm bh, 0}$ = (4.2$^{+1.2}_{-1.0}$) $\times$10$^5\rm M_{\odot} ~ \rm Mpc^{-3}$. All these measurements come from integration of the SMBH mass function by assuming $\epsilon_{\rm rad}=$0.1. 

To fully reconcile the SMBH growth derived here with the observed local BH mass density (e.g. the one from \citealt{Shankar+09}), we need to adopt $\epsilon_{\rm rad}$ = 0.076$^{+0.023}_{-0.018}$, which is in broad agreement with previous measurements. The typical 10 per cent efficiency comes from \citet{Soltan82}, and corroborated by \citet{Fabian+99}. Later on, \citet{Merloni+04} found $0.04 < \epsilon_{\rm rad} < 0.12$ and H07 report $\rho_{\rm bh, 0}$ = 4.81$^{+1.24}_{-0.99} \times $10$^5\rm M_{\odot} ~ \rm Mpc^{-3}$, which results in a slightly higher radiative efficiency ($\epsilon_{\rm rad} \sim $0.1). Finally, MH08 end up with an intermediate value ($\sim$0.07), which is in good agreement with our results.

\section{Conclusions}  \label{conclusions}

We have presented an estimate of the cosmic evolution of SMBH growth, from $z\sim$3 down to the present epoch. Our estimate of the BHAD across the cosmic time is the first one derived by using a far-IR selected sample. We selected 4343 PEP sources with 160$~\mu $m flux density $>$3$\sigma$ detection limit in the GOODS-S and COSMOS fields. The available multi-band photometric coverage allowed us to explore the observed SEDs from the UV to the FIR, in particular thanks to the valuable PACS and SPIRE data from the \textit{Herschel} satellite. Moreover, the availability of \textit{Herschel} data both in the GOODS-S and in COSMOS field was essential to explore different and complementary luminosity regimes. 

Our analysis relies on a robust broad-band SED decomposition performed to decouple the AGN contribution from the host galaxy content in each global SED. We determined reliable estimates of the AGN accretion luminosity, without any bias against highly obscured AGN and based on smaller bolometric corrections than those adopted from X-rays ($\sim$4 instead of 20-30). 

Our main conclusions are as follows.
\begin{itemize}

 \item The percentage of star-forming galaxies in the redshift range $0.1<z<3.8$ harbouring an AGN (with a 99 per cent significance) is as high as 37 per cent.
 \item At $z>0.3$ we find the AGN detection rate rising as a function of the total (1--1000$~\mu $m rest-frame) IR luminosity, suggesting that the AGN activity at these redshifts is more likely to take place in enhanced ($L_{\rm 1-1000} \sim 10^{12} \rm L_{\odot}$) star-forming systems, rather than in ``normal'' ($L_{\rm 1-1000} \sim 10^{11} \rm L_{\odot}$) ones. Furthermore, we find also an additional dependence of the AGN detection rate on the redshift, since the probability to pick up AGN in galaxies with a given luminosity increases with redshift, at least up to $z \sim 2$. At $z<0.3$, the average AGN detection rate reaches 5--10 per cent and does not evolve significantly with IR luminosity. 
 \item We compare our classification based on SED decomposition with that of the \citet{Lacy+07} and \citet{Donley+12} colour-colour AGN selection criteria. As expected, the latter is more reliable and conservative than the former: $\sim$17 per cent of PEP AGN fulfill the \citet{Donley+12} criterion and are characterised by AGN dominated systems in the mid-IR, with extreme ($>10^{12} \rm L_{\odot}$) star formation and AGN accretion luminosities. On the other hand, $\sim 73$ per cent of PEP AGN lie inside the \citet{Lacy+07} wedge: these sources share intermediate properties between QSO-like SEDs and purely star-forming systems, showing more modest IR luminosities $L_{\rm 1-1000} \sim 10^{11-11.5} \rm L_{\odot}$. In addition, we point out that our PEP AGN and PEP galaxies occupy different regions across the mid-IR colour-colour diagram. Such differences highlight the fact that our SED classification effectively labels as ``AGN'' sources with near/mid-IR SEDs systematically different from those of purely SF 
galaxies.
 \item The best-fit AGN bolometric LF is in good agreement with that obtained by H07, despite the AGN selection criteria being completely independent one from the other. 
 \item We find clear evidence in favour of ``downsizing'' behaviour of PEP AGN. Indeed, both from the evolution of the best-fit LF (Fig. \ref{fig:lf_par}, top and middle panel) and from the evolution of the number density with redshift (Fig. \ref{fig:lf_par}, bottom panel), we see that the bulk of the SMBH growth takes place in more powerful objects at earlier times. The similarity between our findings and those previously suggested by X-ray based works (e.g. \citealt{Hasinger+05}; \citealt{Silverman+08}) may imply that AGN hosts, both from X-rays and from FIR selection, have likely experienced similar evolutionary paths. 
 \item The evolution of the BHAD and the resulting BH radiative efficiency ($\epsilon_{\rm rad} \sim 0.07 $) presented in this paper are in agreement with those obtained from X-ray based results. This overall consistency suggests that the FIR selection, together with a proper SED decomposition, allows to obtain similar integrated luminosity densities with respect to X-ray selected AGN samples. However, this does not necessarily mean that the two selection methods trace the evolution of the same AGN population. 
 
\end{itemize}

In this work we have aimed at constraining the mean evolutionary trend of the SMBH growth as seen by \textit{Herschel}-selected galaxies. However, our analysis is not deep enough to shed light on the nature of our AGN sample, or on the evolution of the Compton-thick AGN population. This issue would require principally cross-matching with available X-ray data. In a future work we will investigate more thoroughly the physical properties of the AGN populations by combining IR and X-ray observations, so as to obtain a more complete perspective on the AGN evolution through cosmic time.

\section*{Acknowledgments}

This paper uses data from \textit{Herschel}'s photometers PACS and SPIRE. PACS has been developed by a consortium of institutes led by MPE (Germany) and including: UVIE (Austria); KU Leuven, CSL, IMEC (Belgium); CEA, LAM (France); MPIA (Germany); INAF-IFSI/OAA/OAP/OAT, LENS, SISSA (Italy) and IAC (Spain). This development has been supported by the funding agencies BMVIT (Austria), ESA-PRODEX (Belgium), CEA/CNES (France), DLR (Germany), ASI/INAF (Italy), and CICYT/MCYT (Spain). SPIRE has been developed by a consortium of institutes led by Cardiff Univ. (UK) and including: Univ.Lethbridge (Canada); NAOC (China); CEA, LAM (France); IFSI, Univ. Padua (Italy); IAC (Spain); Stockholm Observatory (Sweden); Imperial College London, RAL, UCL-MSSL, UKATC, Univ. Sussex (UK); and Caltech, JPL, NHSC, Univ. Colorado (USA). This development has been supported by national funding agencies: CSA (Canada); NAOC (China); CEA, CNES, CNRS (France); ASI (Italy); MCINN (Spain); SNSB (Sweden); STFC, UKSA (UK); and NASA (USA).
The authors would like to thank the referee for helpful comments that improved the clarity of the paper. ID is deeply grateful to Iary Davidzon, Andrea Negri, Fabio Vito and Alessandro Marconi for inspiring discussions and suggestions. ID also thanks Elisabeta Lusso for useful tests on codes for SED decomposition and Micol Bolzonella for kindly providing a tool to estimate error bars.

\bibliographystyle{mn2e}
\bibliography{smbh_growth}

\appendix
\section{Incompleteness in accretion luminosity} \label{appendix}

As mentioned in section \ref{monte_carlo}, in this Appendix we describe in detail our approach to account for the incompleteness in accretion luminosity. We follow a similar approach to that developed by \citet{Fontana+04}, who build the stellar mass function correcting for the incompleteness in stellar mass, starting from a \textit{K}-band selected sample. The same argument could be in principle generalised, under some assumptions, to deal with the evaluation of the incompleteness affecting some unobservable quantity when selecting sources with any other observable quantity. We make use of a similar algorithm, but referred to the ``accretion  flux'' $S_{\rm accr}$ as unobservable and to the 160$~\mu$m flux as observable. We define $S_{\rm accr} = L_{\rm bol, AGN} / 4\pi D_{\rm L}^2 $ and compare it with the corresponding 160$~\mu $m flux, simply rescaling the accretion power by its luminosity distance factor $\rm D_{L}$.

As described in section \ref{sample}, the reference sample was selected above a given detection threshold (2.4 mJy in the GOODS-S, 10.2 mJy in the COSMOS field). Nevertheless, we are potentially ruling out some objects that are below the observational limit, but likely having a relatively high accretion power (i.e. a larger $L_{\rm bol, AGN}$ value with respect to sources with FIR flux density above the detection limit). In order to quantify the missing fraction of AGN as a function of the observed 160$~\mu $m flux density, the following hypotheses have been made.

\begin{enumerate}
 \item We assume that the distribution of the flux ratio $S_{\rm accr} / S_{\rm 160}$ does not depend on redshift. Indeed, we are not able to characterise such distribution over sizeable samples at any redshift, expecially at z$>$2 where the low statistics prevents us from deriving the ratio with high significance. 
 
 \item We rely on the basic assumption that the $S_{\rm 160}$ completeness function derived by \citeauthor{Berta+10} (\citeyear{Berta+10}, \citeyear{Berta+11}) for 160$~\mu $m sources in GOODS-S preserves the same trend also below the detection limit. Furthermore, the $S_{\rm 160}$ flux completeness function derived for COSMOS sources has been extrapolated below 10.2 mJy on the existing one, by assuming the same trend for that derived in GOODS-S. 
\end{enumerate}

\begin{figure}
\begin{center}
    \includegraphics[width=\linewidth]{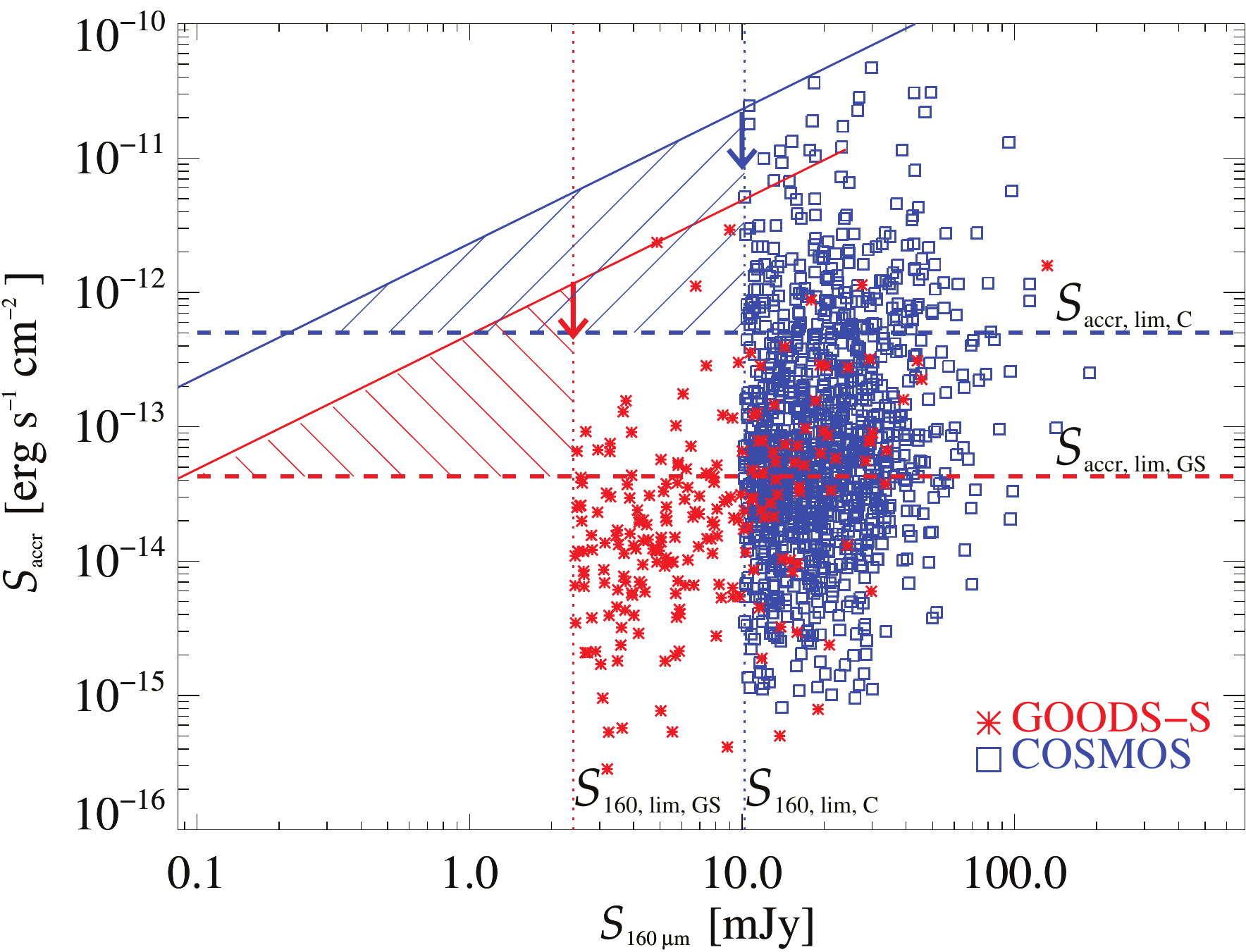}
\end{center}
 \caption{ Distribution of GOODS-S (red asterisks) and COSMOS (blue squares) sources classified as AGN, as a function of $S_{\rm accr}$ and $S_{\rm 160}$. Dotted lines show the flux density limits at 160$~\mu $m ($S_{\rm 160, lim}$), whereas horizontal dashed lines set the thresholds in accretion flux that we found through this analysis. The red and blue solid lines trace the upper value of the distribution of flux ratio $S_{\rm accr}$ / $S_{160}$, as a function of the accretion flux. The red and blue shaded areas define the flux-to-flux regions that we have explored to calibrate our correction for incompleteness in $S_{\rm accr}$. See text for details. }
   \label{fig:completeness}
\end{figure}

We fit the logarithmic distribution of flux ratio with a Gaussian function, for GOODS-S and COSMOS samples separately. In support to the assumption (ii), we observed a similar distribution of flux ratio among COSMOS and GOODS-South, which means that most likely the two fluxes scale in a similar way as a function of $S_{\rm 160}$. Such similar trend at lower $S_{\rm 160}$ allows us to assume that our distribution is also preserved for lower $S_{\rm 160}$ than the GOODS-S detection limit. As expected, the distributions of flux ratio are quite broad (about 0.6 dex), so they do not imply a tight correlation between star-formation and AGN accretion activities (see e.g. \citealt{Rosario+12}). However, the approach developed by \citet{Fontana+04} allows us to take into account the full distribution of flux ratios, even if quite broad. 

In order to correct for the incompleteness in $S_{\rm accr}$, it is necessary to evaluate the intrinsic number of AGN that are missed because $S_{\rm 160}$ is lower than the detection limit ${S_{\rm 160, lim}}$, as a function of the accretion flux. In Fig. \ref{fig:completeness} we plot the distribution of COSMOS (blue squares) and GOODS-S (red asterisks) sources as a function of both 160$~\mu $m and accretion fluxes. The vertical dotted lines set the FIR detection limit of each PEP field. Each coloured solid line marks the upper boundary of the distribution of flux ratios sampled by each value of the accretion flux. For sake of simplicity, hereafter we focus on the red asterisks only (representing GOODS-S). The following reasoning may be also extended to the blue squares (representing COSMOS).

Following \citet{Fontana+04}, we set the starting point for our analysis in such a way that the highest flux ratio of the distribution samples the flux density limit ${S_{\rm 160, lim}}$ value of the survey (see the down-arrow placed where the solid line crosses the dotted one in Fig. \ref{fig:completeness}). We may reasonably assume that above such threshold we are 100 per cent complete in accretion flux. 

Since we are interested in quantifying the missing AGN fraction as a function of the accretion flux, we follow the down-arrow (see Fig. \ref{fig:completeness}), moving along the solid line down to lower and lower values of ${S}_{\rm accr}$.  According to the hypothesis (ii) we assume that also the distribution shifts down with the same trend. For each step in ${S}_{\rm accr}$ we need to evaluate the expected (i.e. intrinsic) number of sources $N_{\rm exp}$, that is nothing more than the ``classic'' $\log N_{\rm exp}$--$\log S_{\rm 160}$. This is achievable by extrapolating the existing $S_{\rm 160}$ completeness curve down to lower and lower FIR flux densities.

As we are interested in deriving such information as a function of the accretion flux rather than FIR flux density, we convert the obtained $\log N_{\rm exp}$--$\log S_{\rm 160}$ into $\log N_{\rm exp}$--$\log S_{\rm accr}$ through a convolution with the distribution of flux ratios. In other words, we use our distribution to derive the expected number of sources for each input accretion flux. From the comparison between the observed and expected number of sources, it is possible to obtain the detectable AGN fraction $f_{\rm det}$ as a function of $S_{\rm accr}$:

\begin{equation}
f_{\rm det} ~ (S_{\rm accr}) = \frac{\mathop{\mathlarger{\int_{S_{\rm 160, lim}}^{S_{\rm 160, max}}{ \frac{dN_{\rm exp}}{dS_{\rm 160}}}}} \cdot g(\overline{x}) ~ dS_{160}}{\mathop{\mathlarger{\int_{S_{\rm 160, min}}^{S_{\rm 160, max}}{ \frac{dN_{\rm exp}}{dS_{\rm 160}}}}} \cdot g(\overline{x}) ~ dS_{\rm 160}}
   \label{eq: f_det}
\end{equation}
where $(\overline{x}) = S_{\rm accr}$ / $S_{\rm 160}$ and $g(\overline{x})$ is the interpolated value of the distribution of flux ratios at [$S_{\rm accr}$, $S_{\rm 160}$]. While $S_{\rm 160, lim}$ is constant, $S_{\rm 160, min}$ and $S_{\rm 160, max}$ change step by step in accretion flux, since they represent the minimum and maximum FIR fluxes, respectively, sampled by the distribution of flux ratios for a given accretion flux $S_{\rm accr}$.

The quantity in equation (\ref{eq: f_det}) is by definition $\leq$1 and scales downward as long as $S_{\rm accr}$ decreases, since the FIR-based selection becomes progressively more incomplete in accretion flux. As done by \citet{Fontana+04}, we choose to run the iterations until the detectable fraction matches the lost fraction (i.e. $f_{\rm det}$ = 0.5). The latter value identifies the minimum accretion flux $S_{\rm accr, lim}$ that a source should have to enter a $>$50 per cent complete accretion-based sample, regardless of its FIR flux density. In Fig. \ref{fig:completeness} the shaded areas delimit the flux-to-flux regions that we have explored to calibrate our correction for incompleteness in accretion flux. 
Such thresholds in accretion flux are equal to $4\times 10^{-14}$ erg s$^{-1}$ cm$^{-2}$ and $5\times 10^{-13}$ erg s$^{-1}$ cm$^{-2}$ in the GOODS-S and the COSMOS field respectively (horizontal dashed lines in Fig. \ref{fig:completeness}).

Within our AGN sample, only those sources with accretion fluxes larger than the corresponding threshold have been considered for the AGN bolometric luminosity function. By applying the above mentioned corrections, the predicted number of sources in each $L_{\rm bol, AGN}$ bin may increase up to a factor of 2. Nevertheless, the $V_{\rm max}$ method (Eq. \ref{eq: vmax}) was referred to the accessible comoving volume for a source to be detected at 160$~\mu $m, but not necessarily to enter at the same time an accretion-based selection. As a consequence, we define an ``effective'' maximum comoving volume $V_{\rm max}^{eff}$ as:
\begin{equation}
V_{\rm max}^{\rm eff} = \min [ V_{\rm max}, V_{\rm max, accr}]
   \label{eq: vmax_eff}
\end{equation}
where $V_{\rm max, accr}$ is the equivalent of the classic $V_{\rm max}$ definition, except for the fact that $V_{\rm max, accr}$ is computed as a function of the accretion flux rather than of FIR flux. In other words, we calculate a $V_{\rm max, accr}$ value for each object by following Eqs. (\ref{eq: vmax}) and (\ref{eq: omega_z}) and replacing the FIR flux completeness function $f_c(z)$ with that based on the accretion flux $f_{\rm det} ~ (S_{\rm accr})$. 

Finally, the effective volume $V_{\rm max}^{\rm eff}$ (Eq. \ref{eq: vmax_eff}) is computed for sources having $S_{\rm accr} \geq S_{\rm accr, lim}$ and $S_{\rm 160} \geq S_{\rm 160, lim}$, fitting both simultaneously with FIR and accretion-based selections. At the same time, $V_{\rm max}^{\rm eff}$ allows us to correct for the incompleteness in accretion flux, that is to place our \textit{Herschel}-selected AGN within the appropriate comoving volume where all AGN with $S_{\rm accr} \geq S_{\rm accr, lim}$, either detected or undetected by \textit{Herschel}, would be observable.

\label{lastpage}

\end{document}